\documentclass{aa}
\usepackage{txfonts}
\usepackage{natbib}
\usepackage{longtable}
\usepackage{graphicx}
\usepackage{threeparttable}
\usepackage{setspace}
\bibpunct{(}{)}{;}{a}{}{,} 

	\title{The barium isotopic fractions in five metal-poor stars}
	\author{A. J. Gallagher\inst{1}
\and
S. G. Ryan\inst{1}
\and
A. Hosford\inst{1}
\and
A. E. Garc\'ia P\'erez\inst{2}
\and
W. Aoki\inst{3}
\and
S. Honda\inst{4}
}
\institute{Centre for Astrophysics Research, School of Physics, Astronomy \& Mathematics, University of Hertfordshire, College Lane, Hatfield, Hertfordshire, AL10 9AB, United Kingdom\\ 
email: \texttt{[a.gallagher, s.g.ryan, a.hosford]@herts.ac.uk} 
\and Department of Astronomy, P.O. Box 400325, University of Virginia, Charlottesville, VA 22904-4325, United States\\
email: \texttt{aeg4x@virginia.edu}
\and National Astronomical Observatory, Mitika, Tokyo, 181-8588, Japan\\
email: \texttt{aoki.wako@nao.ac.jp}
\and  Kwasan Observatory, Kyoto University, Ohmine-cho Kita Kazan, Yamashina-ku, Kyoto, 607-8471, Japan\\
email: \texttt{honda@kwasan.kyoto-u.ac.jp}
}
\date{Received 02 November 2011/ Accepted 29 December 2011}

\authorrunning{A. J. Gallagher et al.}
\titlerunning{The barium isotopic fractions in five metal-poor stars}

\newcommand{\Teff}{T_{\rm eff}}
\newcommand{\kms}{\rm km\,s^{-1}}
\newcommand{\fodd}{f_{\rm odd}}
\newcommand{\foddr}{f_{{\rm odd},r}}
\newcommand{\fodds}{f_{{\rm odd},s}}

\newcommand{\logg}{\log{g}}
\newcommand{\loggf}{\log{gf}}
\newcommand{\SH}{S_{\rm H}}
\newcommand{\Msol}{{\rm M}_{\sun}}
\newcommand{\vconv}{\nu_{\rm conv}}

\newcommand{\ABa}{A(\element{Ba})}
\newcommand{\rt}{\zeta_{\rm RT}}

\abstract{Theory and observations of heavy element nucleosynthesis are in conflict with one-another. Theory states that in the most metal-poor stars, the rapid (r-) neutron-capture nucleosynthetic process would be dominant over the slow (s-) process. The most recent determinations of r- and s-process yields do not support this.}
{We provide measurements of the \element{Ba} isotopic fractions for five metal-poor stars derived with a local thermodynamic equilibrium (LTE) analysis with 1D model stellar atmospheres. This increases the comparisons with heavy element nucleosynthesis theory.}
{We use high resolution ($R\equiv\lambda/\Delta\lambda=90\,000-95\,000$), very high signal-to-noise ($S/N>500$) spectra to determine the fraction of odd \element{Ba} isotopes ($\fodd$) by measuring subtle asymmetries in the profile of the \ion{Ba}{ii} line at 4554\,\AA. We also use two different macroturbulent broadening techniques, Gaussian and radial-tangential, to model the \element{Fe} lines of each star, and propagate each technique to model macroturbulent broadening in the \element{Ba} 4554\,\AA\ line. We conduct a 1D non-LTE (NLTE) treatment of the \element{Fe} lines in the red giant HD\,122563 and the subgiant HD\,140283 in an attempt to improve the fitting. We determine [Ba/Eu] ratios for the two giants in our study, HD\,122563 and HD\,88609, which can also be used to determine the relative contribution of the s- and r-processes to heavy-element nucleosynthesis, for comparison with $\fodd$.}
{We find mathematical solutions of $\fodd$ for HD\,122563, HD\,88609 and HD\,84937 of $-0.12\pm0.07$, $-0.02\pm0.09$, and $-0.05\pm0.11$ respectively. BD$+$26${}^{\circ}$\,3578 yielded a value for $\fodd=0.08\pm0.08$. Only BD$-$04${}^\circ$\,3208  was found to have a physical $\fodd$ ratio of $0.18\pm0.08$. This means that all stars examined here show isotopic fractions more compatible with an s-process dominated composition. The [Ba/Eu] ratios in HD\,122563 and HD\,88609 are found to be $-0.20\pm0.15$ and $-0.47\pm0.15$ respectively, which indicate instead an r-process signature. We report a better statistical fit to the majority of {Fe} profiles in each star when employing a radial-tangential broadening technique during our 1D LTE investigation.}
{With the increase of the number of stars for which the \element{Ba} isotope fraction $\fodd$ has been measured, and the nature of their results, there is now a stronger argument to suggest that other synthesis codes that employ alternative approaches to radiative transfer (e.g. 3D hydrodynamics) have to be considered to tackle the high level of precision required for the determination of isotopic ratios. We have shown that, from a statistical point of view, one must consider using a radial-tangential broadening technique rather than a Gaussian one to model {Fe} line macroturbulences when working in 1D. No improvement to \element{Fe} line fitting is seen when employing a NLTE treatment of the \element{Fe} lines.}

\begin{document}

\maketitle
\section{Introduction}
\label{sec:intro}
Nuclei heavier than the \element{Fe}-peak are mainly synthesised via two neutron-capture processes, the slow (s-) and rapid (r-) process. For the s-process, neutron-capture rates are much lower than $\beta$-decay rates in unstable isotopes, whereas for the r-process, neutron-capture rates are higher than $\beta$-decay rates. Each n-capture process has a different site for nucleosynthesis \citep{B2FH1957}. Low- to intermediate-mass stars ($1\,\Msol\lesssim M\lesssim 8\,\Msol$) evolving along the thermal pulsing asymptotic giant branch (TP-AGB) provide the necessary conditions for the ``main'' s-process, which is responsible for the majority of the s-process elements in the solar-system between $88\leq A\leq204$ \citep{Sneden2008}, releasing free neutrons via $\element[][12]{C}({\rm p},\gamma)\element[][13]{N}(\beta^{+}\nu_{\rm e})\element[][13]{C}(\alpha,{\rm n})\element[][16]{O}$ reaction occurring in the \element{He}-rich zone in radiative conditions \citep{Straniero1997}. Evidence of active s-processing in TP-AGB stars can be seen through absorption line detections of short lived \element[][98]{Tc} and \element[][99]{Tc} isotopes \citep[and references therein]{Smith1988} visible at 4238, 4262 and 4297\,\AA. 

There are many candidates for r-process sites including neutron star explosions \citep{Imshennik1992}, neutron star surface explosions \citep{Bisnovatyi1979} and neutron star winds \citep{Panov2009, Wanajo2001}, to name a few. However, presently the most favoured sites for the r-process are core-collapse supernovae \citep{Wanajo2006}. Extreme temperatures and run away nuclear processes produce an extremely high fluence of free neutrons \citep{Wanajo2003}, which are necessary for r-process nucleosynthesis. 

The presence of seed nuclei with high n-capture cross-sections ($\sigma$), such as \element{Fe}, is critical for n-capture nucleosynthesis. As low- to intermediate-mass stars cannot synthesise nuclides up to the \element{Fe} peak, high-$\sigma$ nuclei must be present at the time of the star's formation for the s-process to occur. High-mass supernova progenitors ($M> 8\Msol$) reach high enough temperatures at the very end of their evolution immediately before the supernova explosive phase to synthesise nuclides up to the \element{Fe} peak. Unlike the s-process, the r-process does not need the presence of high-$\sigma$ nuclei at the time of formation as they are produced in situ shortly before the end of the life of the star. Also, low- to intermediate-mass stars are long lived in comparison to high-mass stars. As such, r-process enrichment from supernovae explosions should dominate in the early universe, i.e. in metal-poor regimes, with the s-process signatures becoming increasingly dominant in more metal-rich regimes. This theory was set out by \citet{Truran1981} and the chemical evolution of n-capture elements from \element{Ba} to \element{Eu} was quantified by \citet{Travaglio1999}. Indications to support this theory can be seen in evidence presented in \citet[their Fig. 14]{Francois2007}.

As the Galaxy becomes more metal-rich over time, s-process signatures in stars begin to increase relative to the r-process for \element{Ba} \citep{Francois2007}. When one compares the [Ba/Eu] ratios from \citet{Francois2007} with those from \citet{Mashonkina2003}, who study the metallicity at which the s-process begins to increase relative to the r-process for the halo and thick disk respectively, there are variations in star-to-star compositions which lead to different r- and s-process regimes for a given metallicity. In particular \citet{Francois2007} find that the s-process begins to increase relative to the r-process at $[{\rm Fe/H}]\gtrsim -2.6$, whereas \citet{Mashonkina2003} find this to occur at $[{\rm Fe/H}]\gtrsim -1.5$.

One potential way of detecting r- and s-process signatures in a star is to measure the isotopic fractions in heavy elements using the profile of their absorption lines. There are differences between pure s- and r-process isotope ratios in most heavy elements, which can be in principle detectable through minute changes in line asymmetry. \element{Ba} is an attractive heavy element for which to use of this method, as the hyperfine splitting (hfs) of its 4554\,\AA\ line from the singly-ionised stage is quite large \citep{Rutten1978} and it offers the possibility of measuring the odd fraction ($\fodd$\footnote{$\fodd = [N(\element[][135]{Ba})+N(\element[][137]{Ba})]/N(\element{Ba})$}) via resolved asymmetric lines.

The r- and s-process are responsible for five of the seven stable isotopes of \element{Ba} (the lightest two, \element[][130,132]{Ba}, arise in the so-called p-process \citep{B2FH1957}). Whereas the s-process can synthesise all five \element{Ba} n-capture isotopes, shielding by \element[][134,136]{Xe} prevents the r-process from synthesising two of the even isotopes, \element[][134,136]{Ba}.

Using nucleosynthesis calculations \citep{Arlandini1999}, the values of $\fodd$ can be determined for the r- and s-process. For a fully s-process regime, $\fodds=0.11\pm0.01$, and in a fully r-process regime, $\foddr=0.46\pm0.06$. \citet{Gallagher2010} show a linear relationship between $\fodd$ and r- and s-process contributions for \element{Ba} determined from \citet{Arlandini1999}. Values of $0.0\leq\fodd<0.11$ or $0.46<\fodd\leq1.0$ are not physical in the context of the nucleosynthesis model but are achievable from more ad hoc isotopic mixes. However, we have assumed that the theory in \citet{Arlandini1999} is accurate and we state throughout our paper that any value of $\fodd$ which lies outside the limits $0.11\leq\fodd\leq0.46$ is non-physical.

From the point of view of spectroscopy, the even \element{Ba} isotopes contribute principally to the formation of the line centre in the \ion{Ba}{ii} 4554\,\AA\ line. The odd isotopes, which are hyperfine split, contribute to the spectral region closer to the wings of the line (see Fig.~\ref{fig:balines}). The relative strength of the odd isotopes located toward the blue wing are smaller than those located toward the red wing and are further from the line core. As such, when the odd isotope contribution to the total line strength is increased, the asymmetry in the absorption line's profile is increased. When \element{Ba} is dominated by the s-process, more of the abundance is associated with the even isotopes, which contribute to the line near its centre; the line profile has a deeper core with shallower wings and the line's asymmetry is reduced.

There are, however, difficulties in determining isotopic fractions in \element{Ba}, chief among which is the high precision required in the analysis of $\fodd$, which demands the highest quality observations and radiative transfer models. Issues then arise when complex astrophysical behaviours, such as convection, start to become visible in the high quality stellar data. Basic assumptions used in conventional spectrum synthesis codes that assume a plane-parallel geometry (1D) and local thermodynamic equilibrium (LTE) cannot actually replicate the observed \ion{Ba}{ii} 4554\,\AA\ line profile \citep{Gallagher2010}. 

The well studied metal-poor subgiant, HD\,140283 illustrates the difficulties that are encountered when determining $\fodd$, which has been attempted several times for this star.  \citet{Gallagher2010} find $[{\rm Fe/H}]=-2.59\pm0.09$ and a low [Ba/Fe] ratio $=-0.87\pm0.14$, which seemingly point to an r-process origin as \element{Ba} is mainly an s-process element in the solar system. Yet according to their isotopic analysis they find $\fodd=0.02\pm0.06$, which agrees with the result published by \citet{Magain1995}, who finds $\fodd=0.08\pm0.06$. Both isotope fractions indicate a fully s-process regime, contradicting the \citet{Truran1981} model. However, measurements of $\fodd$ in 1D LTE by \citet{Lambert2002} and (for the same spectrum) \citet{Collet2009} show HD\,140283 to be slightly r-process dominated with $\fodd=0.30\pm0.21$ and $\fodd=0.33\pm0.13$ respectively, which contradicts \citet{Magain1995} and \citet{Gallagher2010}. When \citet{Collet2009} reanalysed the same spectrum again using 3D hydrodynamic model stellar atmospheres they found $\fodd=0.15\pm0.12$, indicating an s-process signature and supporting results from \citet{Magain1995} and \citet{Gallagher2010}. If the three studies showing HD\,140283 to have an s-process signature are correct, the contradiction with Truran's paradigm could be explained by how s- and r-processes vary with [Fe/H], or by the inhomogeneity of the interstellar medium (ISM) when this halo star formed, in which case star-to-star variations of n-capture signatures would be common. The cause of the discrepancy between the results of the various studies of HD\,140283 is unclear.

Unlike \element{Ba}, \element{Eu} in the solar system has a predominantly r-process contribution, which \citet{Arlandini1999} calculates to be 94\% of the total \element{Eu}. Both stable \element{Eu} isotopes, \element[][151,153]{Eu}, show significant r-process contributions relative to the s-process and these occur in almost equal amounts at a ratio of 0.48:0.52 for 151:153 \citep{Arlandini1999}. High abundances of \element{Eu} in metal-poor stars indicate a strong r-process signature \citep{Spite1978,Sneden2008}. Therefore it is common practice to use the [Ba/Eu] ratio as an indicator of a star's r- and s-process ratio \citep{Burris2000,Honda2006,Honda2007,Sneden2008}. 

Limits on [Ba/Eu] abundance ratios can be set for pure s- and r-processes and are found to be $+1.45$ and $-0.81$ respectively \citep{Burris2000}, which were calculated using solar abundances in \citet{Anders1989}. Using the theoretical abundances in \citet{Arlandini1999}, we find the [Ba/Eu] limits for the s- and r-processes to be $+1.13$ and $-0.69$ respectively.

\citet{Gallagher2010} highlighted the issues that arise when fitting \element{Fe} lines assuming 1D LTE, particularly in the wings of the line, and plotted the average residual for all the \element{Fe} lines analysed. They demonstrated the asymmetries that occur in line formation; 1D LTE radiative transfer codes cannot replicate asymmetries. \citet{Gallagher2010} speculated that a spectroscopic analysis based on 3D radiation-hydrodynamic model stellar atmospheres may resolve these problems. In addition they found irregularities between the observed and synthetic profiles in the \element{Fe} line's core. 

In this paper we determine the isotopic fractions of \element{Ba} in a further five metal-poor stars under the assumption of 1D LTE. We describe the observations in \S\ref{sec:obs}, and the 1D LTE analysis of the \element{Ba} 4554\,\AA\ line in \S\ref{sec:LTE}. Due to the lack of metals in metal-poor stars, i.e. stars with $[{\rm Fe/H}]<-2$, electron number densities are low, which drives down opacities in the atmosphere of late-type stars. Therefore assuming LTE in line forming regions of the stellar atmosphere is no longer valid \citep{Mashonkina2008}. In \S\ref{sec:NLTE} of the work presented here we test whether the irregularities seen in the \element{Fe} line residuals are due to LTE departures, using a non local thermodynamic equilibrium (NLTE) treatment for HD\,140283 and HD\,122563. We discuss the results in \S\ref{sec:discuss}.

\section{Target selection and observations}
\label{sec:obs}
The stellar spectra used in this study are almost the highest quality spectra of very metal-poor stars obtained using the High Dispersion Spectrograph (HDS) \citep{Noguchi2002} at the Subaru 8.2\,m Telescope. All have high resolution ($R\equiv\lambda/\Delta\lambda=90\,000-95\,000$, calculated from the widths of several hundred \element{ThAr} lines), and high signal-to-noise ($S/N>500$ per pixel, as measured around 4500\,\AA). Such spectra are essential for an accurate measurement of $\fodd$. Specifics of the observations can be found in Table~\ref{tab:observations}.

\begin{table}[h]
\begin{center}
\caption{Details of the observations of the stellar spectra.}
\begin{tabular}{l c c c c}
\hline\hline
Star		  &	Date		&	Exp. Time\,(min)& $S/N$	& $R$ 						\\
\hline
HD\,140283	  & 22/07/01	&	82				& 1100	&	$90\,000$				\\
\vspace{-1.5mm}\\
HD\,88609	  & 20/04/04	&	210				& 750 	&	$90\,000$ 				\\
			  & 19/10/05  	&					&	  	&							\\
			  & 20/10/05	&	  				&		&							\\
\vspace{-1.5mm}\\
HD\,122563	  & 30/04/04	&	90				& 850 	&	$90\,000$ 				\\
\vspace{-1.5mm}\\
HD\,84937	  & 22/03/03	&	180				& 630	&	$95\,000$				\\
\vspace{-1.5mm}\\
BD$+$26${}^\circ$\,3578 & 17/05/05	&	130				& 550 	&	$95\,000$				\\
\vspace{-1.5mm}\\
BD$-$04${}^\circ$\,3208 & 18/05/05	&	180				& 580	&	$95\,000$				\\
			  & 19/05/05	&					&	 	&							\\
\hline	
\label{tab:observations}
\end{tabular}
\end{center}
\end{table}

Table~\ref{tab:previousresults} shows previous results on \element{Ba} published for four of the stars studied in this paper. It further illustrates the difficulties in determining $\fodd$, in that for HD\,122563, $\fodd$ and [Ba/Eu] do not support each other\footnote{One might take the view that these results are consistent with a mixed heavy element origin.}, while for HD\,140283 there are large discrepancies between $\fodd$ determinations.

The two giants in our study, HD\,122563 and HD\,88609, and one of the turn-off stars, HD\,84937, show a strong indication that \element{Ba} should be r-process dominated based upon [Ba/Eu] abundance ratios calculated in \citet{Honda2006}, \citet{Honda2007} and \citet{Mashonkina2008}. \citet{Mashonkina2008} complement the [Ba/Eu] determination nicely in HD\,84937 calculating $\fodd =0.43\pm0.14$, indicating an almost fully r-process regime. However they found $\fodd$ in HD\,122563 to be $0.22\pm0.15$, a mostly s-process regime, which contradicts the [Ba/Eu] abundances found in their study and by \citet{Honda2006}, see Table~\ref{tab:previousresults}. We also calculate $\fodd$ for two more turn-off stars, BD$-$04${}^{\circ}$\,3208 and BD$+$26${}^{\circ}$\,3578, neither of which have previous \element{Ba} and \element{Eu} analyses.

\begin{table}[ht]
\begin{center}
\caption{Results from previous 1D LTE studies to determine [Ba/Eu] and/or $\fodd$ for stars studied in this work.}
\begin{tabular}{l c c c c}
\hline\hline
	Star		& 	[Fe/H] 		& 	[Ba/Eu]		& 	$\fodd$			&	Reference	\\
\hline
HD\,140283		&	$-2.59$		&	$>-0.66$	&	$0.01\pm0.04$	&	(1)			\\
				&	$-2.40$		&	$\sim -1.05$&	$0.30\pm0.21$	&	(2)			\\
				&	$-2.50$		&	$\cdots$	&	$0.33\pm0.13$	&	(3)			\\
				&	$-2.70$		&	$\cdots$	&	$0.08\pm0.06$	&	(4)			\\
HD\,122563		&	$-2.77$		&	$-0.50$		&	$\cdots$		&	(5)			\\
				&	$-2.53$		&	$-0.41$		&	$0.22\pm0.15$	&	(6)			\\
HD\,84937		&	$-2.15$		&	$-0.69$		&	$0.43\pm0.14$	&	(6)			\\
HD\,88609		&	$-3.07$		&	$-0.48$		&	$\cdots$		&	(7)			\\
\hline
\label{tab:previousresults}
\end{tabular}
\end{center}
(1) \citet{Gallagher2010}, as measured by the 4554\,\AA\ line.
(2) \citet{Lambert2002}.
(3) \citet{Collet2009}.
(4)	\citet{Magain1995}.
(5)	\citet{Honda2006}.
(6) \citet{Mashonkina2008}.
(7)	\citet{Honda2007}.
\end{table}

\section{1D LTE analysis}
\label{sec:LTE}
In this section we briefly review the method used to determine the isotopic fractions, abundances and broadening values of the stars in our sample. All synthetic spectra in this section were created using the {\tiny ATLAS} \citep{Cottrell1978} radiative transfer code with {\tiny KURUCZ06} model atmospheres (\texttt{http://kurucz.harvard.edu/grids.html
}). For a more extensive description of the processes involved in the following procedure, we refer the reader to \citet{Gallagher2010}. Results are provided in Table~\ref{tab:1Dresults}.

\subsection{Barium line lists}
\label{sec:ba}
Line lists with all hfs components of the \element{Ba} line at 4554\,\AA, which is used to determine $\fodd$, were constructed using isotopic information from \citet{Arlandini1999} and hfs information from \citet{Wendt1984} and \citet{Villemoes1993} for pure s- and pure r-process mixtures (corresponding to $\fodd=0.11$ and $0.46$ respectively). Hybrid line lists for $-0.24\leq\fodd\leq 0.46$ were created from these by adjusting the line strengths of the \element{Ba} isotopes. Further details on this can be found in \citet{Gallagher2010} but we remind the reader that $\fodd=0.11$ is the lowest value of $\fodd$ achieved in the s-process of \citet{Arlandini1999}; $\fodd=0.00$ is the lowest value of $\fodd$ achievable in an even-only isotope mix, and that values of $\fodd<0.00$ are non-physical, mathematical solutions only. 

Isotopic abundances presented in \citet{Arlandini1999} which we use to determine $\foddr$ and $\fodds$ were normalised \citep[by][]{Arlandini1999} to the s-process-only isotope \element[][150]{Sm}. We explored a renormalisation to \element[][134]{Ba} and \element[][136]{Ba}, but found little to no change in $\foddr$: 0.49 (renormalised to \element[][134]{Ba}) and 0.46 (renormalised to \element[][136]{Ba}). The renormalisation does not affect $\fodds$, which remains $\fodds=0.11$. This does not significantly alter the interpretation of the results presented here.

No attempt was made to determine $\fodd$ for the 4934\,\AA\ line because, as showed by \citet{Gallagher2010}, analysis of this line is extremely difficult and yields large errorbars due to \element{Fe} blends found in the wings of this line. Nor do we attempt to study higher excitation lines of \element{Ba}, as their hyperfine splitting is much smaller than that at 4554\,\AA.

\subsection{Determination of the macroturbulence}
\label{sec:macro} 
To recap on \citet{Gallagher2010}, we use a $\chi^{2}$ code (derived from that of \citet{Garcia2009}) to compare the observed and synthetic spectra of a number of \ion{Fe}{i} and \ion{Fe}{ii} lines, computed for different abundances $A(\element{Fe})$\footnote{$A({\rm X})={\rm log}_{10}\left(\frac{N({\rm X})}{N({\rm H})}\right)+12$.}, wavelength shifts $\Delta\lambda$, and line broadening parameters. Two different broadening approaches are used, as described below. We derive, for each \element{Fe} lines analysed, the best fitting abundance, wavelength shift and broadening. An ordinary least squares (OLS) fit is calculated through the wavelength-dependence of broadening and $A(\element{Fe})$ values to determine the best values at 4554\,\AA.

\subsubsection{Gaussian broadening}
\label{sec:Gaussian}
One approach to the macroscopic broadening is to adopt a Gaussian of ${\rm FWHM}=\vconv$, representing the convolution of a Gaussian instrumental profile with a Gaussian macroturbulent profile. As \citet{Lambert2002} show, $\fodd$ is extremely sensitive to $\vconv$. They found $\delta\fodd/\delta\vconv=-0.51\,(\kms)^{-1}$ for HD\,140283. \citet{Gallagher2010} found an even larger sensitivity, $-0.71\,(\kms)^{-1}$. The effects of this large sensitivity can be reduced by increasing the number of \element{Fe} lines, $N$, used to constrain $\vconv$, as we take the error in $\vconv$ as the standard error, $\sigma/\sqrt{N}$, where $\sigma$ is the standard deviation of $\vconv$ values. In our work, only \element{Fe} lines of comparable equivalent widths ($W$) to the \ion{Ba}{ii} line are selected, so they would have similar formation depths to the \element{Ba} line, implying that macroturbulent effects on \element{Ba} would be well described by \element{Fe}. We also do not analyse strong lines where uncertain pressure effects in the line broadening become significant.

The adopted atmospheric parameters for each star are listed in Table~\ref{tab:1Dresults} (rows (1) to (5)) and the derived broadening is given in row (7).

\begin{figure*}[hp]
\begin{center}
	\resizebox{0.49\hsize}{0.32\vsize}{\includegraphics{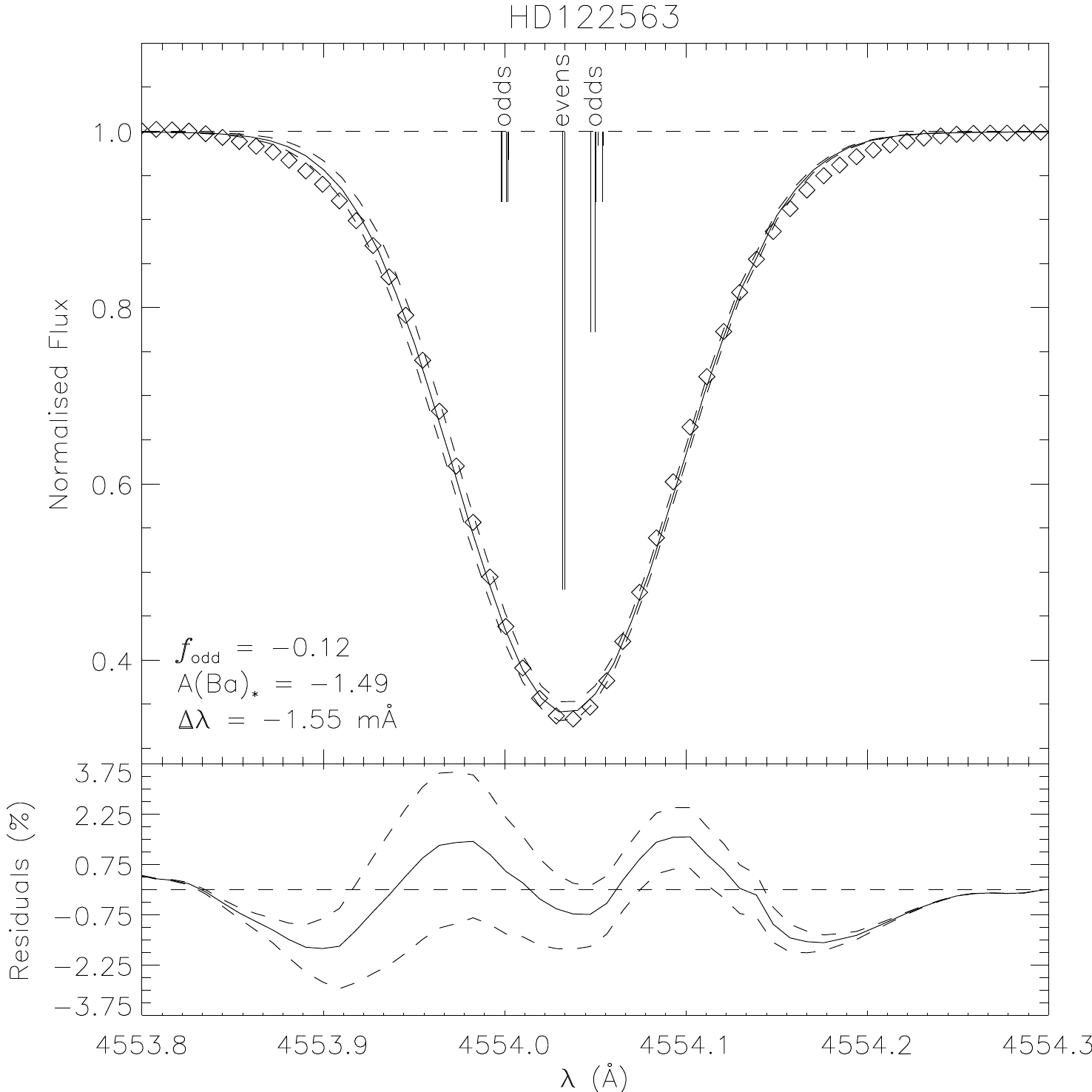}}
	\resizebox{0.49\hsize}{0.32\vsize}{\includegraphics{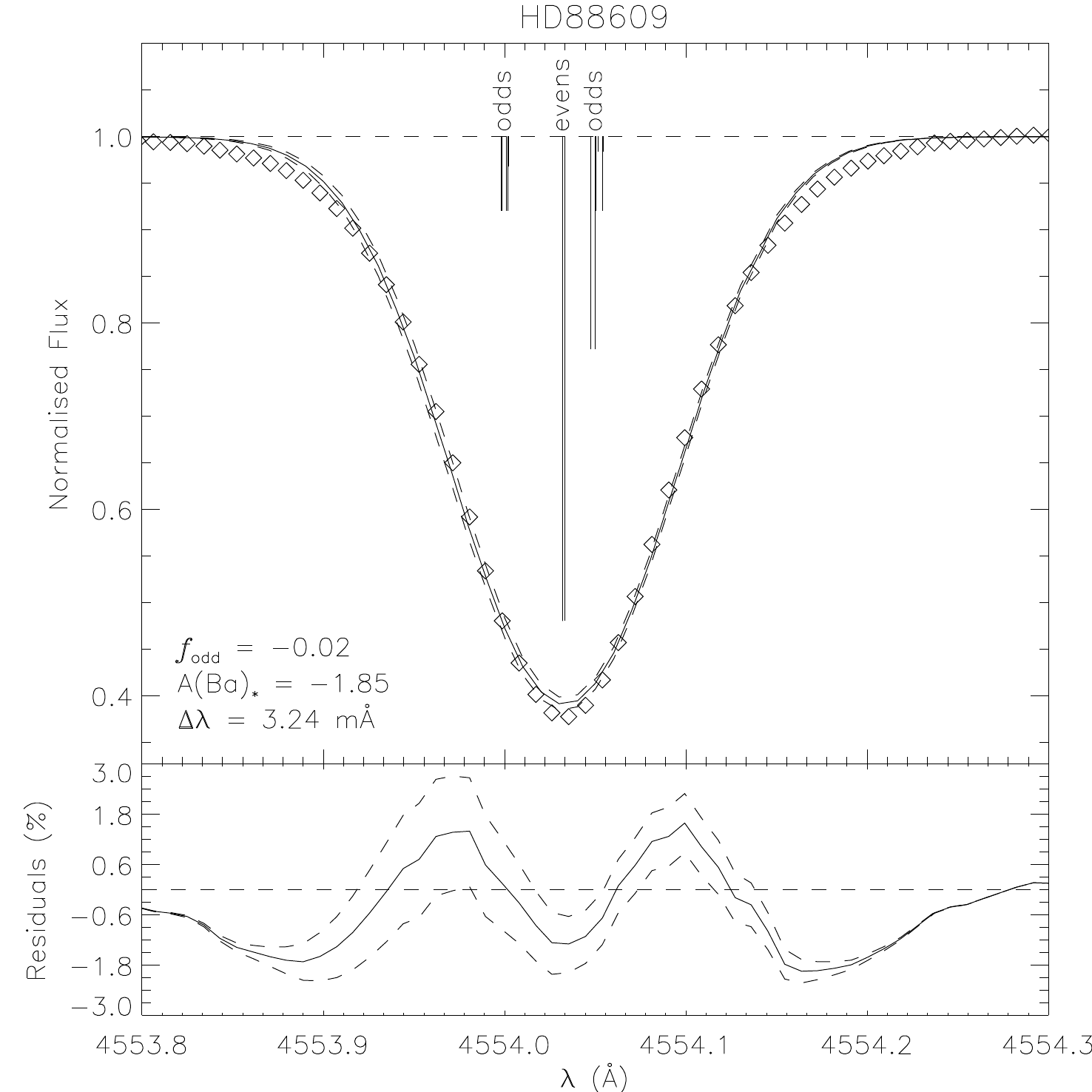}}
	\resizebox{0.49\hsize}{0.32\vsize}{\includegraphics{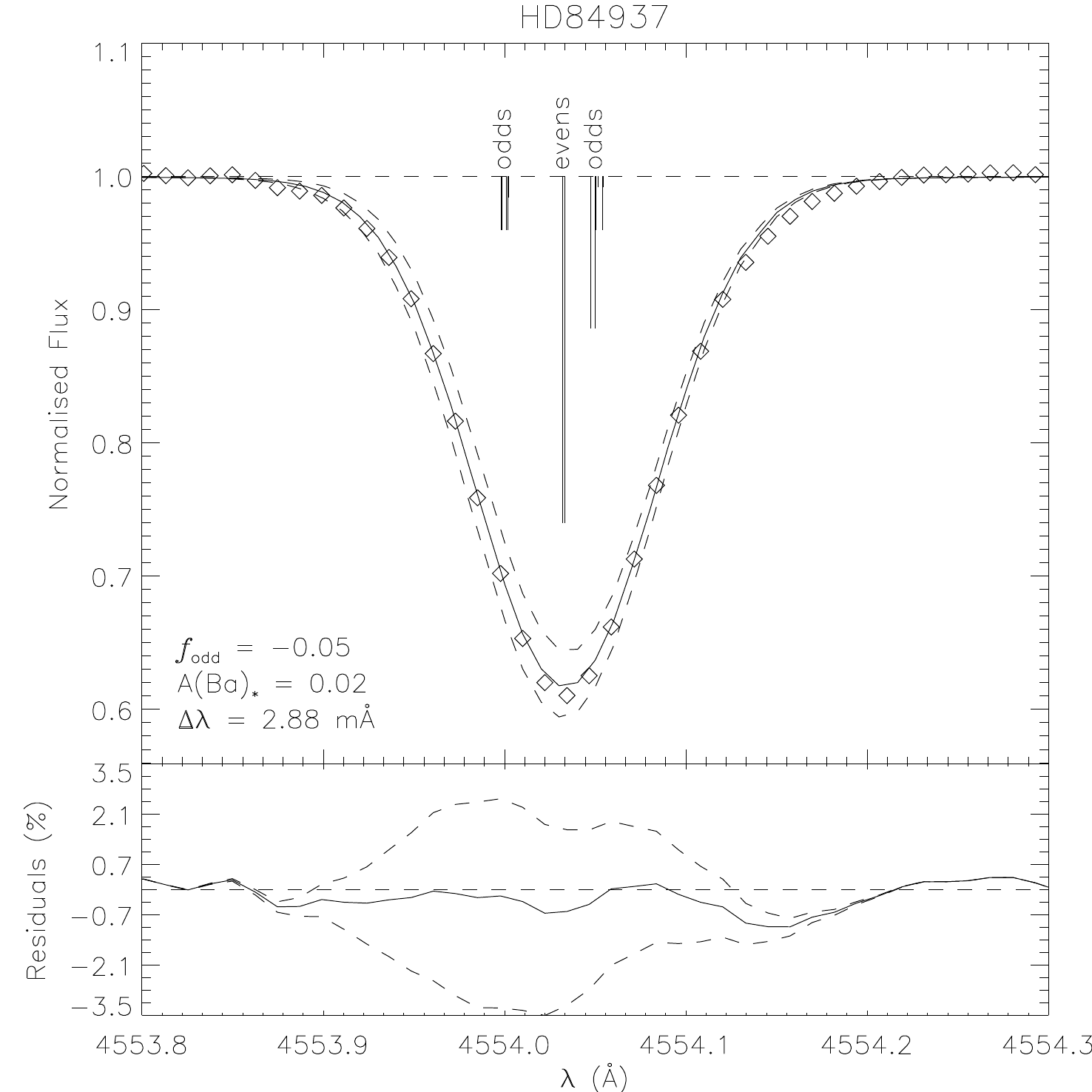}}
	\resizebox{0.49\hsize}{0.32\vsize}{\includegraphics{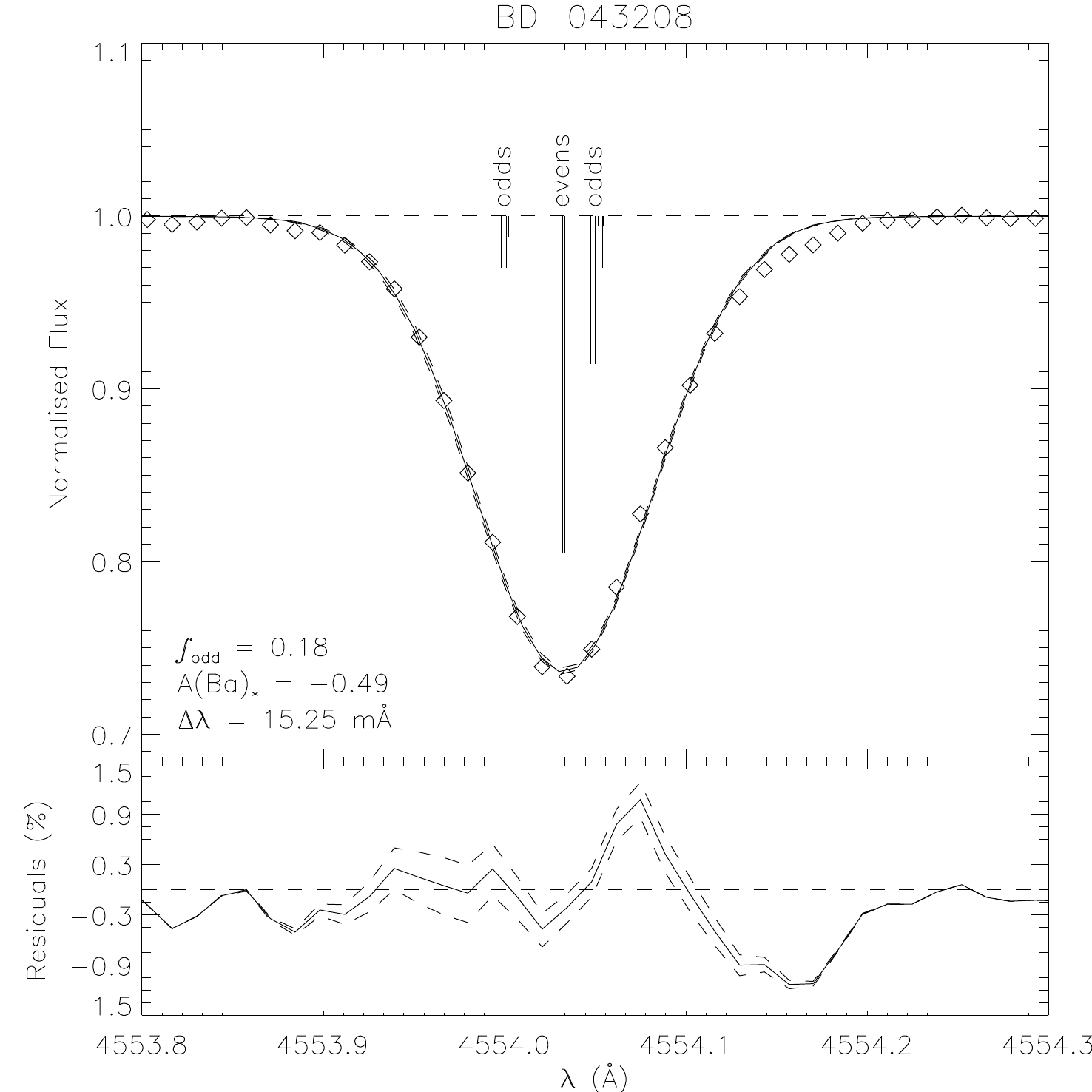}}
	\resizebox{0.49\hsize}{0.32\vsize}{\includegraphics{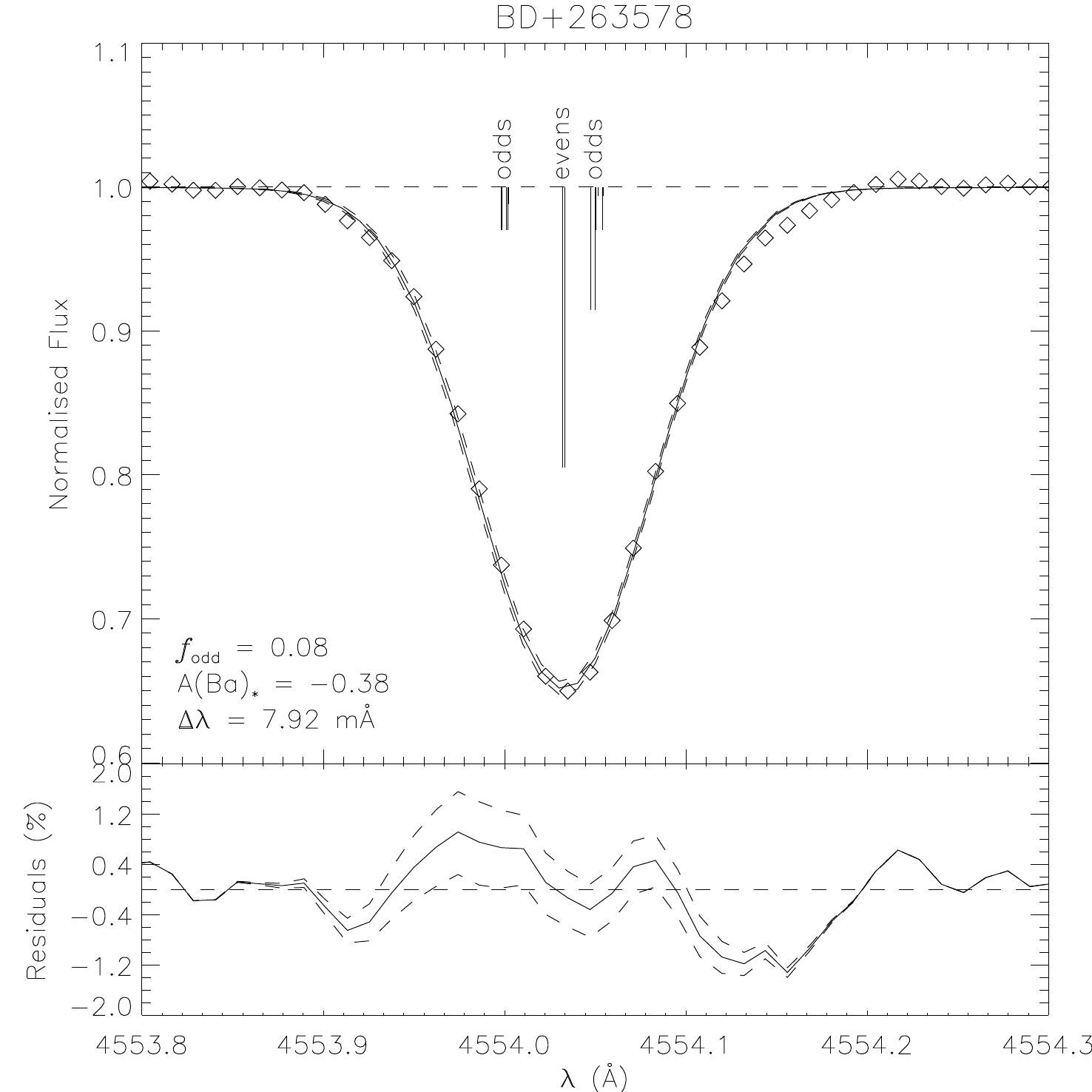}}
	\resizebox{0.49\hsize}{0.32\vsize}{\includegraphics{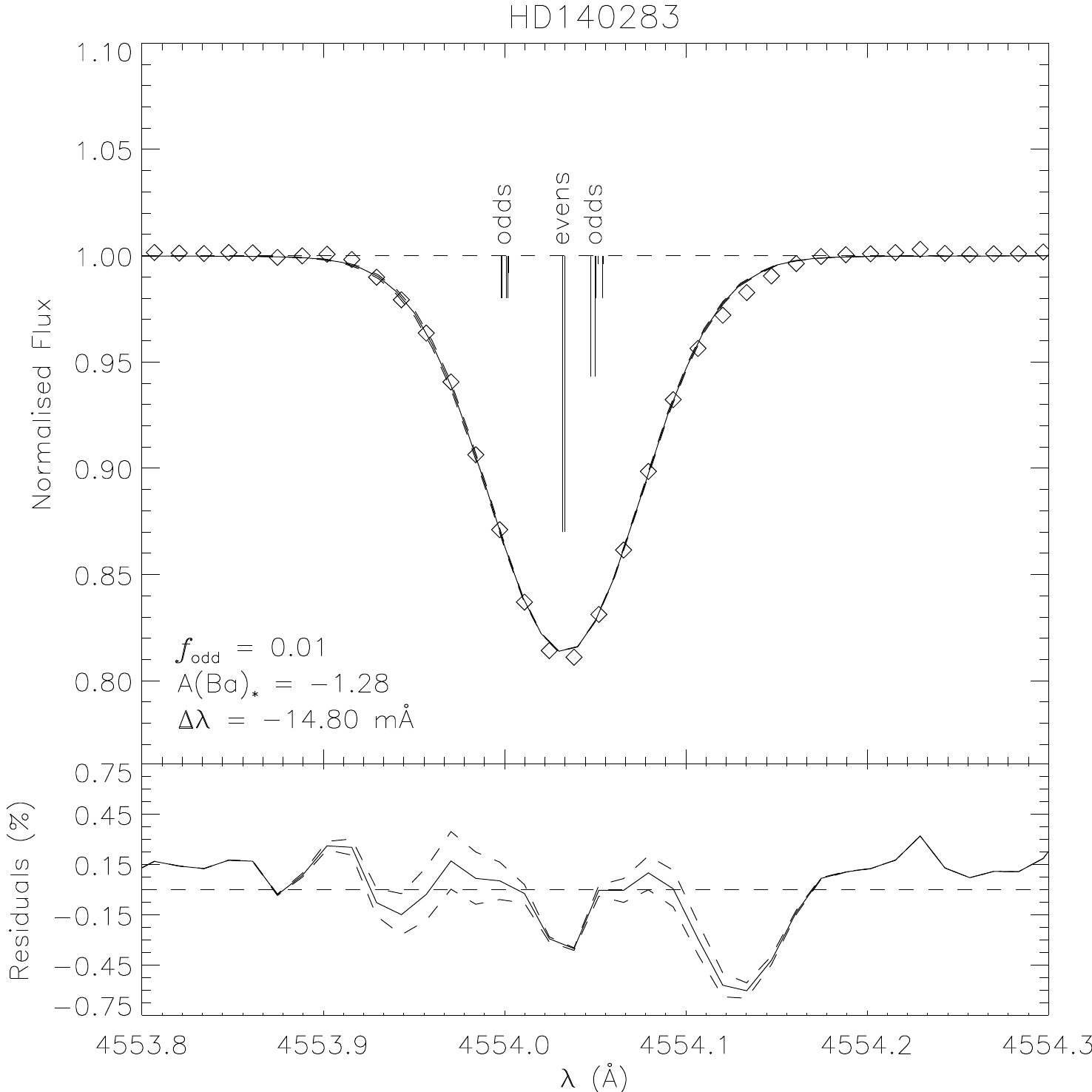}}
	\caption{The best fit \ion{Ba}{ii} 4554\,\AA\ lines for each star, using a Gaussian broadening technique. Each figure displays the observed \element{Ba} profile (diamonds) and the best fit synthetic profile (solid line), which includes the error on $\fodd$ (dashed line). We have also included a schematic of the odd and even isotopes for reference. The lower panel of each figure shows the residuals (obs-syn) of each fit as a percentage. For reference we have included HD\,140283, which was analysed in \citet{Gallagher2010}.}
	\label{fig:balines}
\end{center}
\end{figure*}
\begin{figure*}[hp]
\begin{center}
	\resizebox{0.49\hsize}{0.32\vsize}{\includegraphics{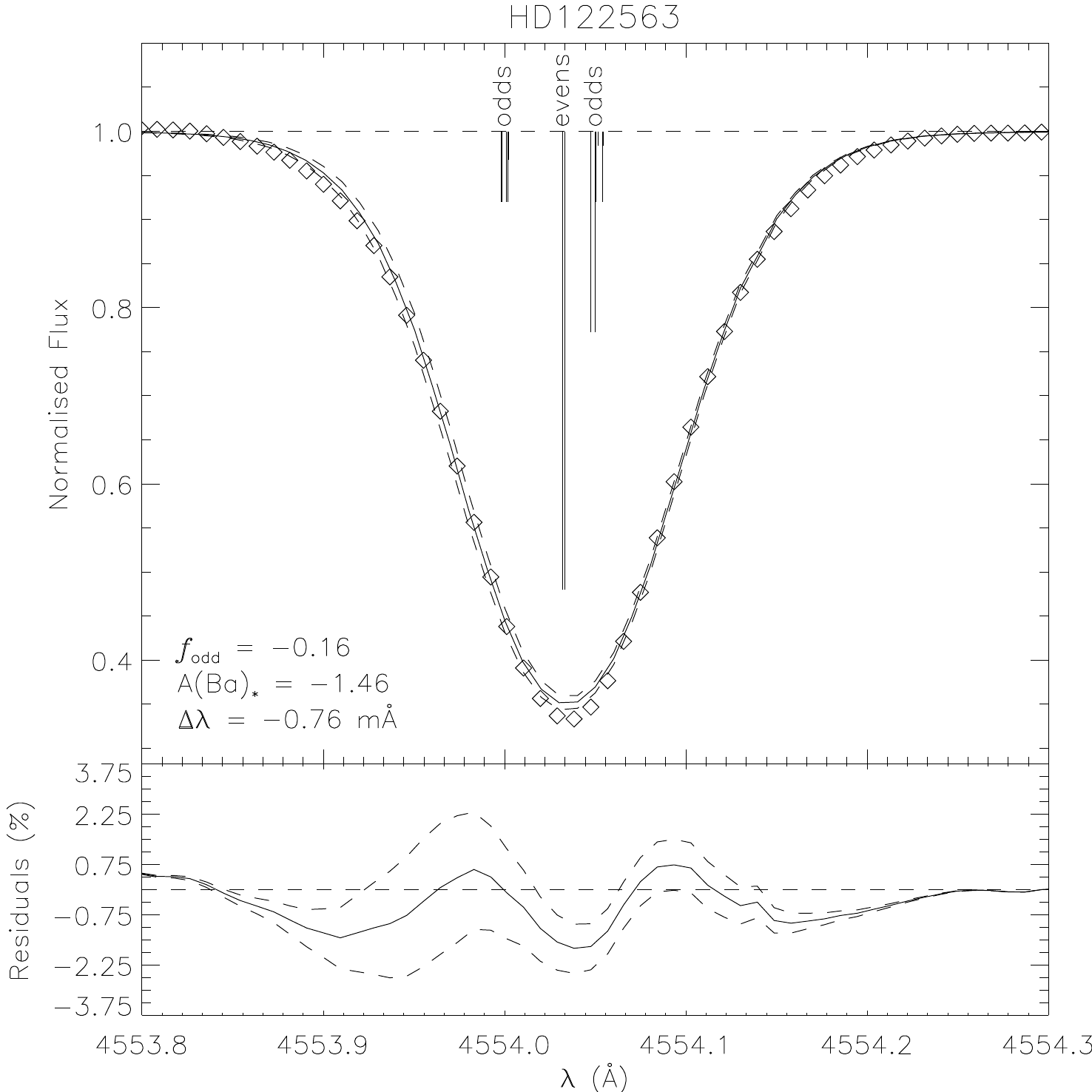}}
	\resizebox{0.49\hsize}{0.32\vsize}{\includegraphics{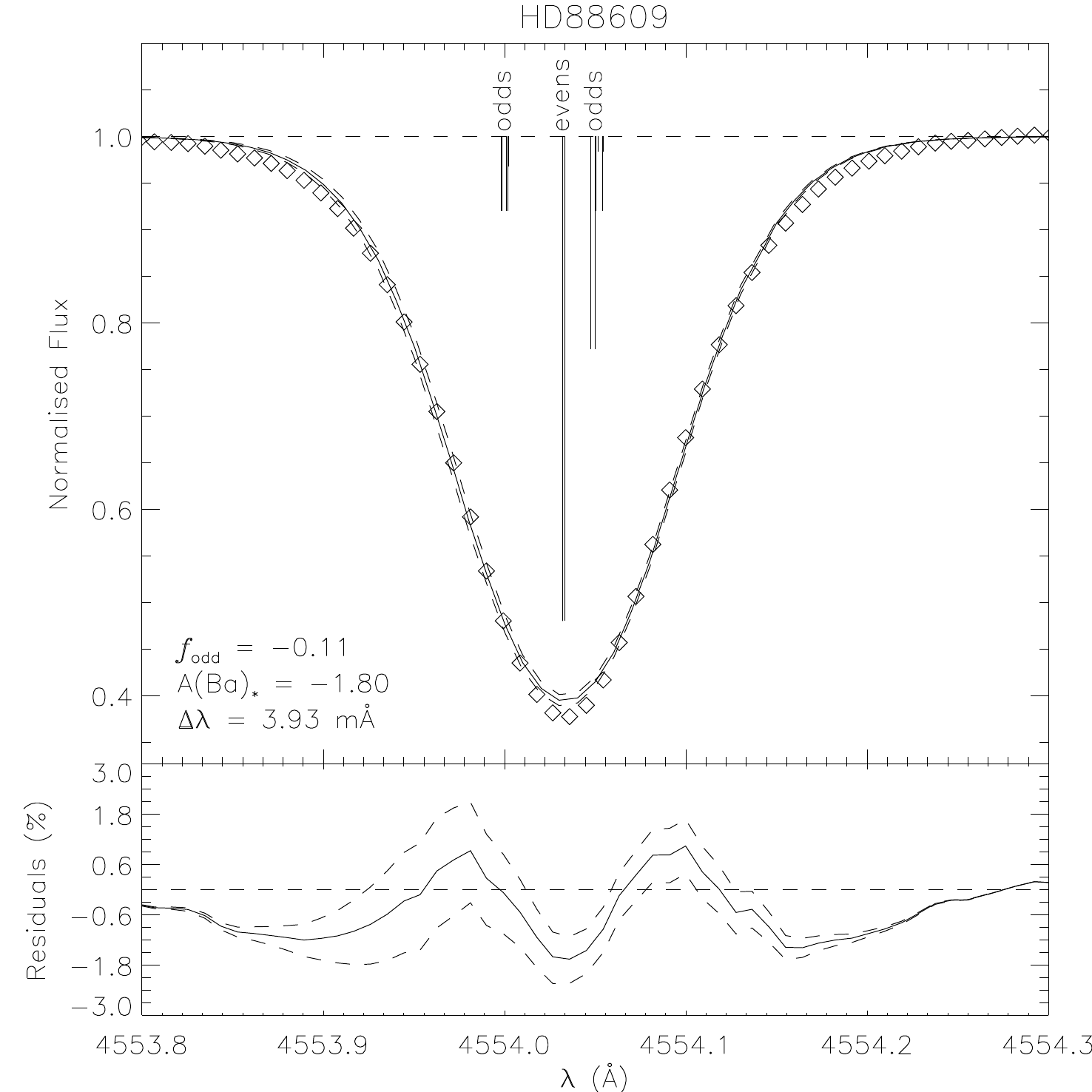}}
	\resizebox{0.49\hsize}{0.32\vsize}{\includegraphics{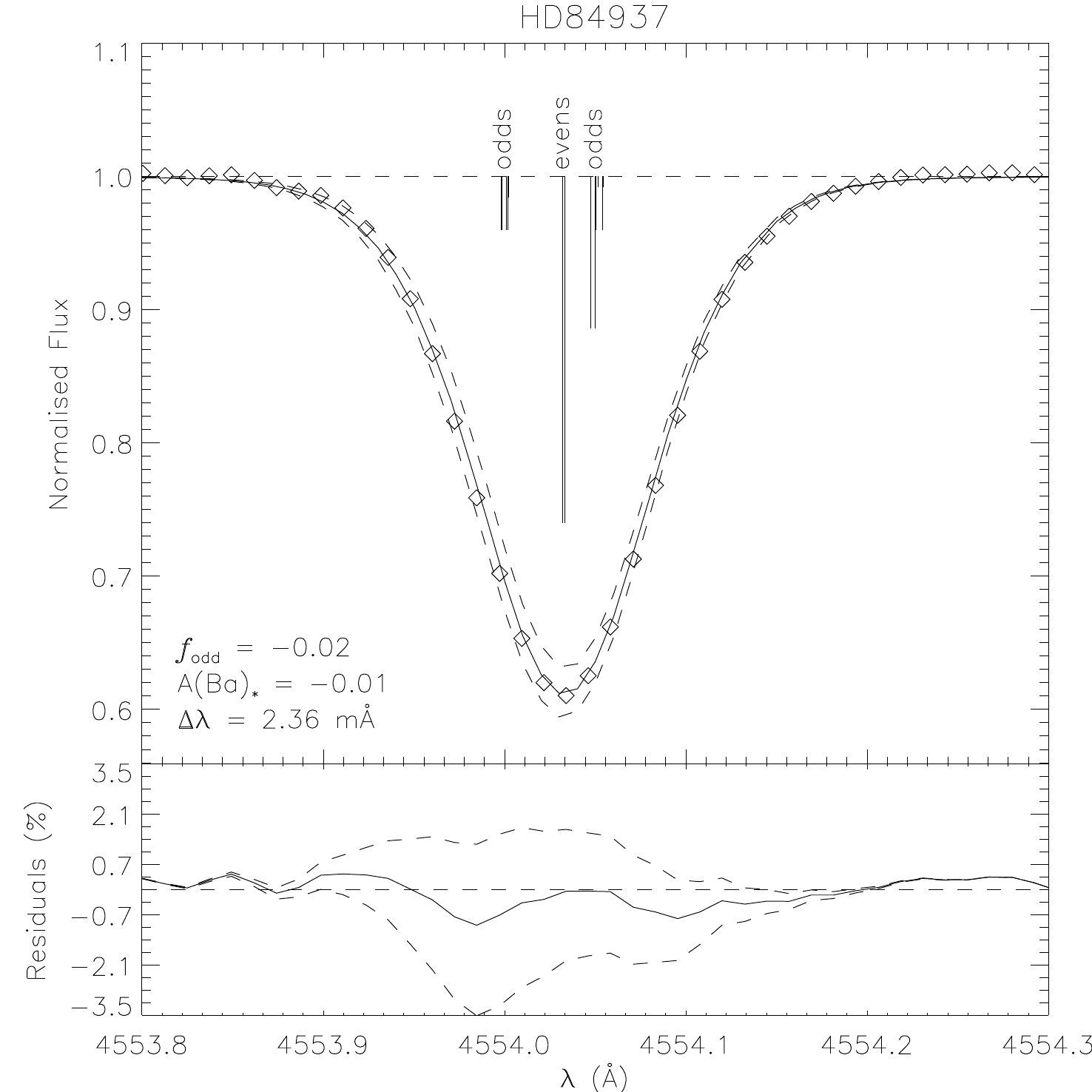}}
	\resizebox{0.49\hsize}{0.32\vsize}{\includegraphics{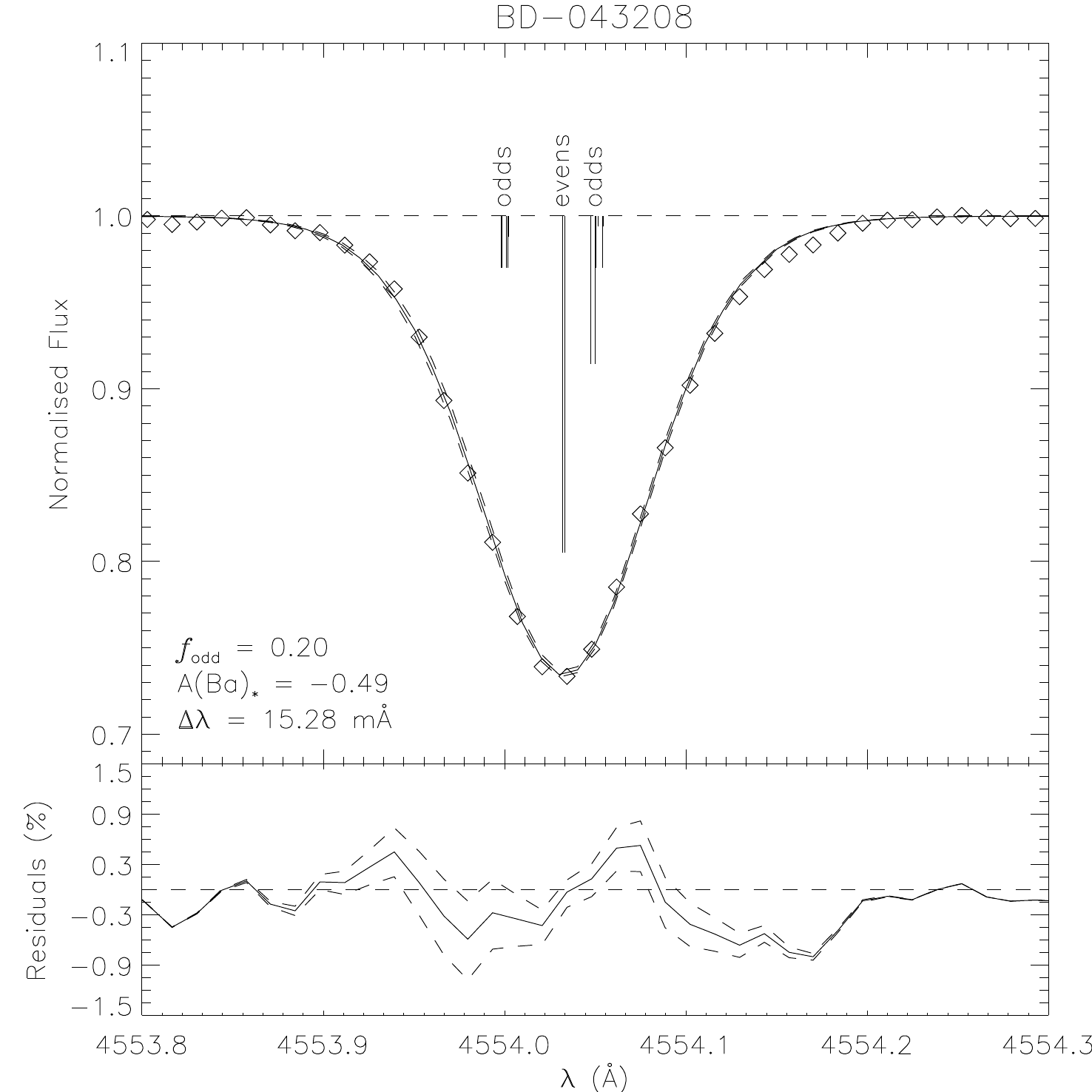}}
	\resizebox{0.49\hsize}{0.32\vsize}{\includegraphics{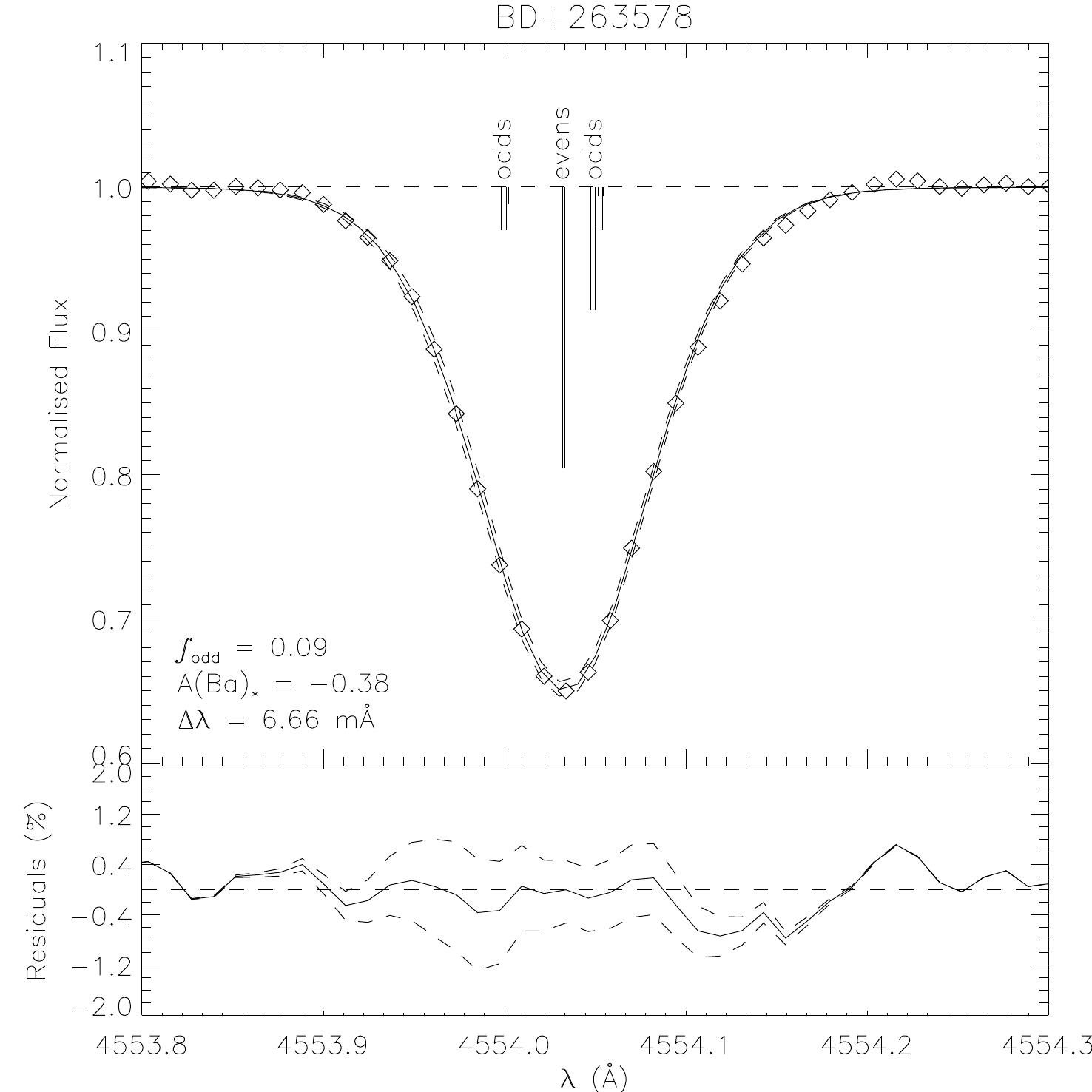}}
	\resizebox{0.49\hsize}{0.32\vsize}{\includegraphics{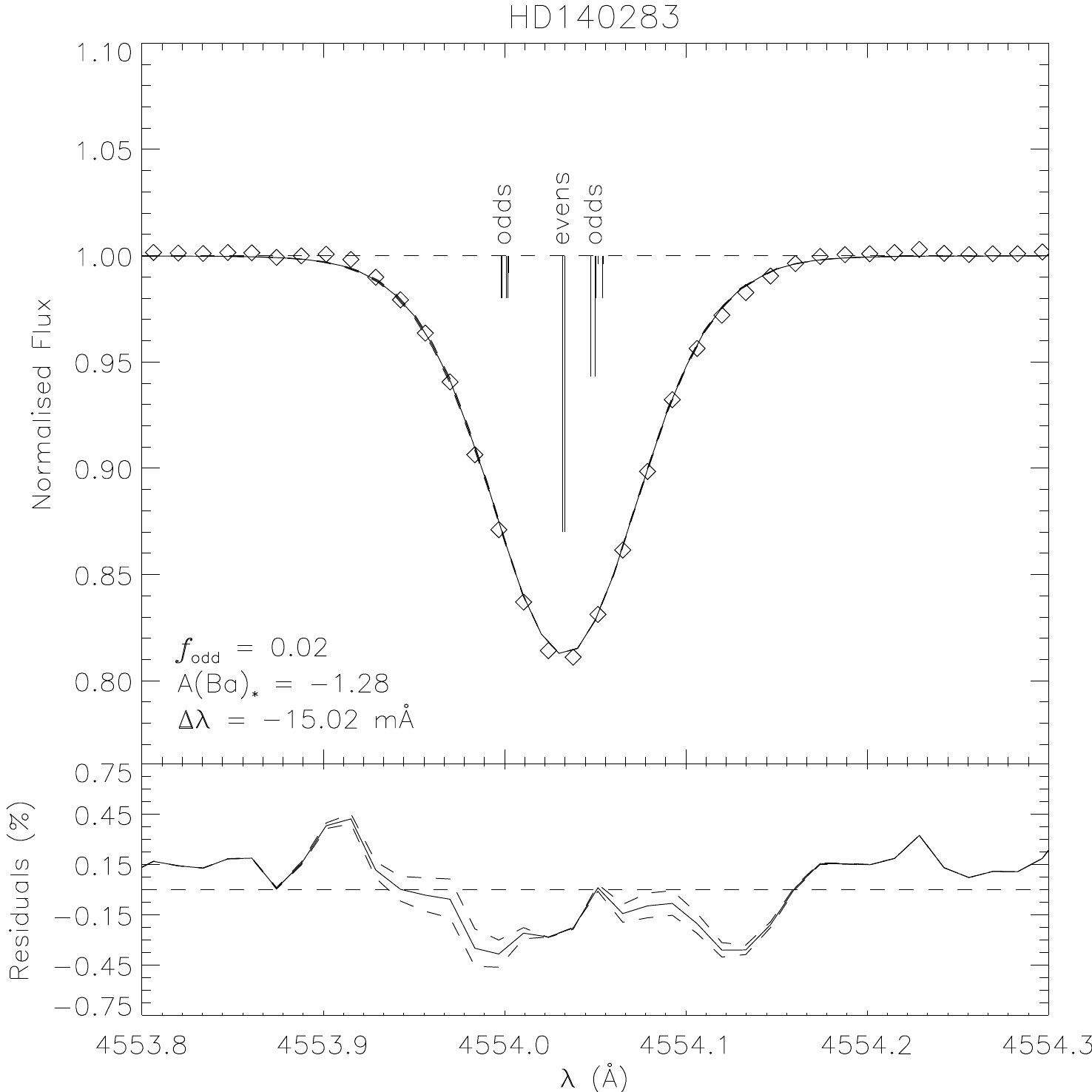}}
	\caption{The best fit \ion{Ba}{ii} 4554\,\AA\ lines for each star, using a radial-tangential broadening technique. Each figure displays the observed \element{Ba} profile (diamonds) and the best fit synthetic profile (solid line), which includes the error on $\fodd$ (dashed line). We have also included a schematic of the odd and even isotopes for reference. The lower panel of each figure shows the residuals (obs-syn) of each fit as a percentage.}
	\label{fig:balinesrt}
\end{center}
\end{figure*}

\begin{table*}[tp]
\begin{center}
\caption{Results of the 1D LTE analysis conducted on five metal-poor stars and the various parameters used in their analysis.}
\begin{tabular}{@{}l l@{} r r r r r r@{}}
\hline\hline
&Parameter										&	HD\,122563	&	HD\,88609	&	HD\,84937	&	BD$-$04${}^{\circ}$\,3208	&	BD$+$26${}^{\circ}$\,3578	& HD\,140283 	\\
\hline
\multicolumn{8}{c}{\textit{Atmospheric Parameters}} \\
\hline
(1)&$\Teff \ ({\rm K})$									& $4570\pm100$	& $4550\pm100$	& $6290\pm100$	& 	$6340\pm100$	& 	$6240\pm100$	& $5750\pm100$	\\
(2)&$\logg \ ({\rm cm\,s^{-1}})$						& $1.1\pm0.3$	& $1.1\pm0.3$	& $3.9\pm0.3$	& 	$4.0\pm0.2$		& 	$3.9\pm0.2$		& $3.7\pm0.1$	\\
(3)&$[\element{Fe}/\element{H}]$						& $-2.77\pm0.19$& $-3.07\pm0.20$& $-2.15\pm0.30$& 	$-2.28\pm0.20$	& 	$-2.33\pm0.20$	& $-2.50\pm0.20$\\
(4)&$\xi \ (\kms)$										& $2.2\pm0.5$	& $2.4\pm0.5$	& $1.2\pm0.3$	& 	$1.5\pm0.2$		& 	$1.5\pm0.2$		& $1.4\pm0.1$	\\
(5)&Atmosphere Reference								& (1)			& (2) 			& (3) 			& (4) 				& (4)				& (5)			\\
\hline
\multicolumn{8}{c}{\textit{Gaussian broadening}} \\
\hline
(6)&$\fodd$												& $-0.12\pm0.07$& $-0.02\pm0.09$& $-0.05\pm0.11$& $0.18\pm0.08$		& 	$0.08\pm0.08$	& $0.02\pm0.06$	\\
(7)&$\vconv \ (\kms)$									& $6.99\pm0.07$	& $7.03\pm0.08$	& $6.98\pm0.06$	&  $6.62\pm0.05$	& 	$6.41\pm0.06$	& $5.75\pm0.02$	\\
(8)&$[\element{Fe}/\element{H}]$						&$-2.90\pm0.17$	&$-3.17\pm0.17$	& $-2.24\pm0.11$&  $-2.42\pm0.11$	& 	$-2.45\pm0.12$	& $-2.59\pm0.09$\\
(9)&$[\element{Ba}/\element{Fe}]$						&$-0.85\pm0.21$	& $-0.91\pm0.21$& $0.04\pm0.15$	&  $-0.20\pm0.15$	& 	$-0.06\pm0.16$	& $-0.87\pm0.14$\\
(10)&$[\element{Ba}/\element{Eu}]$						& $-0.20\pm0.15$& $-0.47\pm0.15$& 	$\cdots$	& 	$\cdots$		& 	$\cdots$		&	$>-0.66$	\\
(11)&$\langle\Delta\lambda\rangle \ ({\rm m\AA})$		&$-15.86\pm0.37$&$-14.70\pm0.90$&$-14.69\pm0.53$& $-11.47\pm1.093$	&	$-13.29\pm0.89$	&$-12.67\pm0.48$\\
\vspace{-1.5mm}\\
(12)&$\delta\fodd/\delta\vconv \ (\kms)^{-1}$			& 	$-0.57$		& 	$-0.74$		&   $-0.89$	    &     $-0.78$	 	&	  $-0.77$	 	&    $-0.71$	\\
(13)&$\delta\fodd/\delta\xi \ (\kms)^{-1}$				&	$-0.08$		&	$-0.08$		&	$0.05$		&	  $0.05$		&	  $0.05$		&	 $0.00$		\\
(14)&$\delta\fodd/\delta\Teff \ (100 \ {\rm K})^{-1}$	&	$-0.04$		&	$-0.05$		&	$-0.03$		& 	  $-0.02$		& 	  $-0.02$		&	 $-0.02$	\\
(15)&$\delta\fodd/\delta\logg \ ({\rm dex})^{-1}$		&	$0.02$		&	$0.02$		&	$-0.27$		&	  $-0.23$		&	  $-0.23$		&	 $-0.28$	\\
(16)&$\delta\fodd/\delta E_\gamma \ (0.7)^{-1}$			&	$-0.02$		&	$-0.02$		&	$-0.05$		&	  $-0.04$		&	  $-0.04$		&	 $\cdots$	\\
\hline
\multicolumn{8}{c}{\textit{Radial Tangential broadening}} \\
\hline
(17)&$\fodd$											& $-0.16\pm0.05$& $-0.11\pm0.07$& $-0.02\pm0.10$& $0.20\pm0.10$		& 	$0.09\pm0.10$	& $0.02\pm0.04$ \\
(18)&$\zeta_{\rm RT} \ (\kms)$							& $5.69\pm0.09$	& $5.74\pm0.09$	& $5.77\pm0.08$	& $5.36\pm0.06$		& 	$5.14\pm0.07$	& $4.30\pm0.02$	\\
(19)&$\nu_{\rm inst} \ (\kms)$							&	$3.32$		&   $3.32$		&   $3.20$		&		$3.20$		&		$3.20$		&	 $3.32$		\\
(20)&{\footnotesize number of best fit \element{Fe} lines}	&	53			&		34		&		30		&		28			&		32			&		38		\\
\vspace{-1.5mm}\\
(21)&$\delta\fodd/\delta\zeta_{\rm RT} \ (\kms)^{-1}$	& 	$-0.30$		& 	$-0.64$		&   $-0.60$	    &     $-0.73$	 	&	  $-0.70$	 	&    $-0.63$	\\
(22)&$\delta\fodd/\delta\xi \ (\kms)^{-1}$				&	$-0.08$		&	$-0.08$		&	$0.05$		&	  $0.05$		&	  $0.05$		&	 $0.00$		\\
(23)&$\delta\fodd/\delta\Teff \ (100 \ {\rm K})^{-1}$	&	$-0.01$		&	$-0.001$	&	$-0.05$		& 	  $-0.07$		& 	  $-0.06$		&	 $0.01$		\\
(24)&$\delta\fodd/\delta\logg \ ({\rm dex})^{-1}$		&	$-0.01$		&	$-0.02$		&	$-0.18$		&	  $-0.22$		&	  $-0.21$		&	 $-0.35$	\\
\hline
\multicolumn{8}{c}{\textit{Other spectral information}} \\
\hline
(25)&Spectral range\,(\AA) 								&$3080$ - $4780$&$3070$ - $4780$&$4130$ - $6860$& $4130$ - $5340$	& $4050$ - $5250$	& $4100$ - $6900$\\
(26)&\element{Fe} line sample size						& 54			& 	35			&		44		&		36			&		37			&	  93		\\
(27)& $W_{\element{Ba}} \ ({\rm m}\AA)$					&  99.1			&  93.2			&   49.7		&   34.7			&   42.7			&   20.1		\\
\hline
\label{tab:1Dresults}
\end{tabular}
\end{center}
(1) \citet{Honda2006}.
(2) \citet{Honda2007}.
(3) \citet{Aoki2009}.
(4) \citet{Garcia2009}.
(5) \citet{Gallagher2010}.
\end{table*}

\subsubsection{Radial-tangential macroturbulent broadening}
\label{sec:rt}
In \citet{Gallagher2010} we compared three different broadening techniques to find out which of them best fit the \element{Fe} lines. We found that using a rotational broadening mechanism, where the macroturbulent broadening of the star was represented by $\nu\sin{i}$ only (with the instrumental broadening still represented by a Gaussian), almost always gave a worse fit than when we employed a simple Gaussian. However we found that using a radial-tangential macroturbulent broadening technique ($\rt$) allowed us to fit spectral lines slightly better than the Gaussian mechanism. As a result, we have continued to employ this broadening type in the current investigation for all five stars and again for HD\,140283, as well as the simple Gaussian approach. The prescription we adopt for $\rt$ assumes equal speeds in the radial and tangential flows, and equal temperatures \citep{Graybook}.

The model spectra were synthesised with {\tiny ATLAS} using a grid of $\rt$ values ranging from $4.00\,\kms$ to $9.00\,\kms$ in steps of $0.2\,\kms$. The instrumental broadening, determined from \element{ThAr} lines, and represented as a Gaussian, was also included in the synthesis. Each line was fit using the same $\chi^2$ code that was used in \S\ref{sec:Gaussian}. Results for $\rt$ can be found in Table~\ref{tab:1Dresults} (rows (17) to (24)), where we have also listed the number of lines that are best fit by the $\rt$ approach for each star.

\subsection{The isotopic fraction of barium}
\label{sec:fodd}
Once $\vconv$ and $\rt$ were obtained for the two broadening mechanisms, the \ion{Ba}{ii} 4554\,\AA\ line was analysed using a $\chi^{2}$ code very similar to the one described in \S\ref{sec:macro} to find values for the free parameters $\Delta\lambda$, $\ABa$ and $\fodd$ that minimise $\chi^{2}$ for the 4554\,\AA\ line in each star. Results for $\fodd$ for Gaussian and radial-tangential broadening techniques can be found in Table~\ref{tab:1Dresults}. The best fit \element{Ba} profiles are shown in Fig.~\ref{fig:balines} (Gaussian) and in Fig.~\ref{fig:balinesrt} (radial-tangential). From the Gaussian results we find for HD\,122563, HD\,88609, HD\,84937, BD$-$04${}^\circ$\,3208 and BD$+$26${}^{\circ}$\,3578 that $\fodd=-0.12\pm0.07$, $-0.02\pm0.09$, $-0.05\pm0.11$, $0.18\pm0.08$ and $0.08\pm0.08$ respectively. These results would suggest that all stars examined here show a high s-process fraction.

\subsection{The [Ba/Eu] ratio}
\label{sec:[Ba/Eu]}
In our sample only the two giants had \element{Eu} line strengths adequate to conduct a \element{Eu} abundance analysis. This was done by examining the \ion{Eu}{ii} 4129\,\AA\ line. The \ion{Eu}{ii} 4205\,\AA\ line is not used in this study as it is blended with a \ion{V}{ii} line \citep{Honda2006,Gallagher2010} that can affect abundance determinations. Also no attempt is made to analyse the isotopic splitting of \element[][151,153]{Eu}; we assume a fixed 50:50 isotopic split of 151:153 when constructing the hfs-affected \element{Eu} 4129\,\AA\ line list. The \element{Eu} line list for the solar r- and s-process ratio was constructed using hfs information from \citet{Becker1993} and \citet{Krebs1960} with $\loggf$ values taken from \citet{Biemont1982}. Abundances for \element{Eu} are taken as those that satisfy the $\chi^{2}$ minimum. From these abundances, and those found by the \element{Ba} analysis, [Ba/Eu] was calculated and the results for both stars can be found in Table~\ref{tab:1Dresults}, row (10). 

\subsection{Error analysis}
\label{sec:errors}
In \citet{Gallagher2010} we demonstrated how altering one stellar parameter can force other parameters, in particular $\vconv$, to compensate for its effect on $\fodd$. In that paper we called this case 1. The compensation by $\vconv$ was well illustrated when one looked at the effects of changing the microturbulence ($\xi$). Once $\vconv$ was recalculated from the \element{Fe} lines, the effect $\xi$ had on $\fodd$ was nullified. As $\vconv$ partially compensates for other changes in the stellar parameters as well, i.e. $\Teff$ and $\logg$, their effect on $\fodd$ is also reduced. 

To calculate the error in $\fodd$ we look at five possible sources of error: $\vconv$, $\xi$, $\Teff$, $\logg$ and the Uns{\"o}ld approximation enhancement factor, $E_{\gamma}$, which enhances the effect of $\gamma_{6,{\rm vdW}}$ in the Van Der Waals calculation, $\gamma_{6}=\gamma_{6,{\rm vdW}}E_\gamma$. In our analysis we have used $E_{\gamma}=2.2$. To test the effect it has on $\vconv$ and $\fodd$ we have decreased this to 1.5. The effect of uncertainties in [Fe/H] on $\fodd$ is negligible and as such was ignored \citep{Gallagher2010}. It is expected that every star belonging to the same stage of evolution, i.e. giant, sub-giant and turn-off, would reproduce comparable sensitivities to each stellar parameter. As such we have run each test for two of the five stars: to test the sensitivity of $\fodd$ in the giants we have used HD\,88609, and for the turn-off stars we have chosen BD$-$04${}^{\circ}$\,3208. However, we have run sensitivity tests of $\vconv$ and $\rt$ for all stars as these parameters have the largest effect on $\fodd$. The results of the sensitivity of $\fodd$ can be seen in Table~\ref{tab:1Dresults}, rows (12) to (16) and rows (21) to (24). The tests confirm that the $\fodd$ determinations are essentially unaffected by the choice of $E_\gamma$, $\Teff$ or $\xi$, but $\logg$ and the macroturbulent broadening effects are more important. It has been reported by \citet{Tajitsu2010} that the EEV 42-80 CCDs used in the HDS suffer from nonlinearity. By investigating the differences between the corrected and non-corrected spectra, we found that its effect on $\fodd$ was negligible.

\section{1D NLTE {Fe} line analysis}
\label{sec:NLTE}
In \citet{Gallagher2010} we aimed to constrain the macroturbulence of HD\,140283 through the use of the \element{Fe} lines. Using the same method, we have also done this for five further stars in \S\ref{sec:LTE}. This was done under the assumptions of LTE. However, it is well known that \ion{Fe}{i} suffers from the effects of NLTE \citep{Thevenin1999,Shchukina2005} in metal-poor stars. We therefore sought to quantify NLTE effects for \element{Fe} on the preceding analysis, in particular on the determination of macroturbulence, and therefore the value of $\fodd$. No attempt is made to determine NLTE corrections for \element{Ba} itself, either in $\fodd$ or the [Ba/Eu] ratio. \citet{Mashonkina2008} demonstrated that corrections to the abundance ratio are not significant enough to change the inferred r- and s-process regime in a star. The determination of $\fodd$ assuming NLTE goes beyond the scope of the work presented in this section.

To compute the \ion{Fe}{i} profiles, a version of the NLTE code {\tiny MULTI} \citep{Carlsson1986} was used with modifications to include the effects of line-blanketing, described in \citet{Collet2005}. The code employs {\tiny MARCS} model atmospheres. The model atom used was that adopted by \citet{Hosford2010}. For a longer discussion on the model atom, processes of NLTE radiative transfer and its impact on stars of this type, see \citet{Hosford2010}. It is important, however, to mention the parameter $\SH$, the scaling factor for the collisions due to \element{H} as described by the approximate Drawin formula \citep{Drawin1968,Drawin1969}. Due to the uncertainties in the magnitude of the \element{H} collisions, the value of $\SH$ is still uncertain and is treated differently by different works. \citet{Collet2005} treats it as a free parameter and tests values $\SH = 1$ and $0.001$. \citet{Korn2003} found a higher value, $\SH = 3$, however in more recent work \citep{Mashonkina2010} the same group has constrained it to $0.1$ based on an improved \element{Fe} atom. Here we adopt the values 1 and 0.001; this gives us two sets of synthetic spectra, one with the Drawin ($\SH=1$) description of \element{H} collisions, and a second close to maximal NLTE effects ($\SH=0.001$). We also compute line profiles in LTE using {\tiny MULTI}.

NLTE calculations of \ion{Fe}{i} line profiles were performed for two stars, the subgiant HD\,140283 and the giant HD\,122563. The first was analysed so a comparison could be made between this work and the LTE analysis of \citet{Gallagher2010}. The second was analysed to understand the NLTE effects on \ion{Fe}{i} in giants, where the atmospheres are more tenuous and also cooler, to act as a comparison to the LTE work in this paper.

Three sets of {\tiny MULTI} runs where completed for each star, one in LTE, and two in NLTE with  the $\SH$ values mentioned above. A $\chi^{2}$ analysis procedure similar to the one described in \S\ref{sec:LTE} been used. Synthetic profiles were created over the parameter space shown in Table~\ref{tab:nlteparam}. The extrinsic broadening was represented by a Gaussian consolidating both macroturbulence and instrumental broadening, the values ${\rm FWHM}_{\rm G}$ in Table~\ref{tab:nlteparam} representing the FWHM of the Gaussian.
	
\begin{table}[h]
\begin{center}
\caption{Ranges of parameters used to create the synthetic profiles in {\footnotesize MULTI}.}
\begin{tabular}{@{} l l c c c @{}}
\hline\hline
&& \multicolumn{3}{c}{$\chi^2$ grid parameter ranges}\\
\cline{3-5}\\
\vspace{-6mm}\\
& & $A(\element{Fe})$& ${\rm FWHM}_{\rm G}$ & $\Delta\lambda$ \\
Star&Run&&(m\AA)&(m\AA)\\
\hline
HD\,140283 	& 	${\rm LTE}_{\rm MULTI}$	&	$4.28-5.28$	&	$73-119$ 	&	$\pm25$	\\
			&	$\SH =1$				&	$4.66-5.28$	&	$73-119$ 	&	$\pm25$	\\
			&	$\SH =0.001$			&	$5.00-5.70$	&	$73-119$ 	&	$\pm25$	\\
\vspace{-2.5mm}\\
HD\,122563	&	${\rm LTE}_{\rm MULTI}$	&	$4.00-5.58$	&	$74-114$ 	&	$\pm25$	\\
			&	$\SH =1$				&	$4.00-5.58$	&	$74-114$ 	&	$\pm25$	\\
			&	$\SH =0.001$			&	$4.68-5.50$	&	$74-114$ 	&	$\pm25$	\\
\hline
\label{tab:nlteparam}
\end{tabular}																											
\end{center}																											
\end{table}

The intention was to see if the observed \ion{Fe}{i} profiles could be better fit by spectra computed in NLTE, in particular for the giants, where the cooler atmospheres result in larger line strengths (at fixed [Fe/H]). Here, the synthetic LTE core is too shallow and the wings too broad even at the minimum $\chi^{2}$ value of the line. In switching to NLTE, we hoped we might model the \element{Fe} lines better, and hence obtain better values of macroturbulent broadening and, in later works, a more reliable value of $\fodd$ for \element{Ba}.

Table~\ref{tab:nltelte} gives the results from the $\chi^{2}$ analysis using {\tiny MULTI}, and the {\tiny ATLAS} LTE results for comparison. The {\tiny MULTI} values are the mean of 51 lines for HD\,140283 and 31 lines for HD\,122563; errors are the standard error. Fewer lines are used in the {\tiny MULTI} analysis than in the {\tiny ATLAS} LTE analysis due to the incompleteness of the adopted model atom which leads to several of the lines not being computable in {\tiny MULTI}. Also, after further scrutinizing the results, three lines from HD\,140283 and three lines from HD\,122563 were removed from the analysis. This was due to either a poorly defined continuum, or blending lines very close to the line of study.

\begin{table}[h]
\begin{center}
\caption{Results from the NLTE analysis and the ATLAS LTE results for comparison.}
\begin{tabular}{@{} l l c c @{}}
\hline\hline\vspace{-3.2mm}\\
Star & Run & $\langle A(\element{Fe})\rangle$ & $\langle\vconv\rangle\,(\kms)$ \\
\hline
HD\,140283 	& 	${\rm LTE}_{\rm MULTI}$	&	$4.96\pm0.01$	&	$6.03\pm0.04$ \\
			&	$\SH =1$				&	$5.28\pm0.01$	&	$5.98\pm0.05$ \\
			&	$\SH =0.001$			&	$5.53\pm0.01$	&	$5.92\pm0.04$ \\
			&	${\rm LTE}_{\rm ATLAS}$	&	$4.92\pm0.04$	&	$5.76\pm0.09$ \\
			&	(51 line subset)		&					&				  \\
			&	${\rm LTE}_{\rm ATLAS}$	&	$4.91\pm0.01$	&	$5.75\pm0.02$ \\
			&	(93 lines)				&					&				  \\
\vspace{-2.5mm}\\
HD\,122563	&	${\rm LTE}_{\rm MULTI}$	&	$4.58\pm0.03$	&	$7.02\pm0.10$ \\
			&	$\SH =1$				&	$4.98\pm0.03$	&	$6.95\pm0.09$ \\
			&	$\SH =0.001$			&	$5.12\pm0.03$	&	$6.91\pm0.09$ \\
			&	${\rm LTE}_{\rm ATLAS}$	&	$4.54\pm0.11$	&	$7.06\pm0.18$ \\
			&	(31 line subset)		&					&				  \\
			&	${\rm LTE}_{\rm ATLAS}$	&	$4.51\pm0.04$	&	$6.99\pm0.07$ \\
			&	(54 lines)				&					&				  \\

\hline
\label{tab:nltelte}
\end{tabular}																											
\end{center}																											
\end{table}																											

\begin{figure*}[t]
\begin{center}
	\resizebox{1\hsize}{!}{\includegraphics{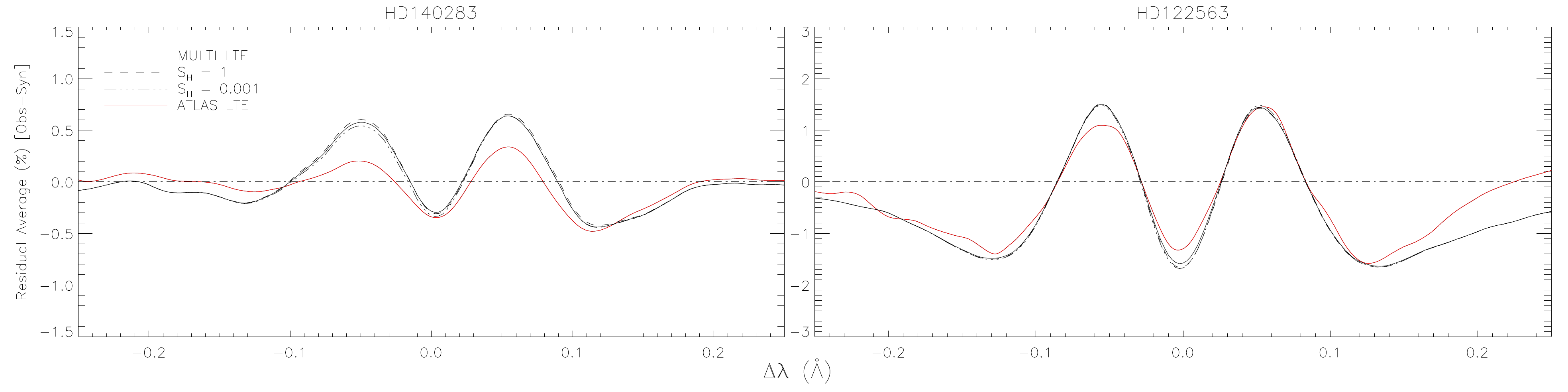}}
	\caption{Plots of average residuals from $\chi^{2}$ analysis of \element{Fe} lines in HD\,140283 and HD\,122563, for $\SH =1$, 0.001 and LTE using {\tiny MULTI}. For comparison we have generated the same plot using {\tiny ATLAS} output data for the same subsets of lines.}
	\label{fig:NLTELTEresidualcompare}
\end{center}
\end{figure*}

A comparison of {\tiny MULTI} LTE and {\tiny ATLAS} LTE results shows that the abundances and extrinsic broadening value are comparable for both cases, the only exception being the broadening values of HD\,140283, which is the most important parameter as it directly affects $\fodd$. For HD\,140283 we have larger values in NLTE than that of \citet{Gallagher2010}. Table~\ref{tab:nltelte} shows that for HD\,140283, choosing the same 51 line subset doesn't ``fix'' the discrepancy. For HD\,122563 the difference is much smaller and is within the errors. One could naively ascribe this difference to be discrepancies in the {\tiny MARCS} and {\tiny KURUCZ} model atmospheres, e.g. different temperature gradients and density gradients. However, a comparison between the {\tiny MARCS} and {\tiny KURUCZ} model atmospheres can be found in \citet{Adamthesis}, who found little to no effect by these parameters. It is most likely that the differences arise from the different treatments of the broadening itself in {\tiny MULTI} and {\tiny ATLAS}.  We see comparable wavelength shifts between the two analyses, and comparable profiles for each line. In NLTE we see an increase of abundance with decreasing $\SH$ due to the overionisation effects becoming more pertinent; thus a larger positive abundance is needed for neutral lines to compensate. As a consistency check, NLTE abundances from the $\chi^{2}$ analysis were compared to those from an equivalent width analysis and were found to be comparable at each $\SH$ value.

An interesting thing to note here is the sensitivity of $\vconv$ to the $\SH$ value and the difference between LTE and NLTE values. For both stars, $\vconv$ decreases by $0.1\,\kms$ in going from {\tiny MULTI} LTE to {\tiny MULTI} $\SH =0.001$. Naively, one might infer from Table~\ref{tab:1Dresults} that this would increase $\fodd$ from an LTE analysis of \element{Ba} 4554\,\AA\ between 0.06 and 0.09. However, some of this difference may not fully apply to \element{Ba} calculated in LTE, due to different broadening under NLTE. Nonetheless, it points to the importance of an accurate description of the radiative transfer if one is to be able to determine $\fodd$.

One hope of the NLTE analysis was that it may aid in fitting the stronger \element{Fe} lines of the giant by producing more realistic line profiles, and hence reduce the residuals in the \element{Fe} analysis \citep[see][their Fig. 10]{Gallagher2010}. In Fig.~\ref{fig:NLTELTEresidualcompare} we regenerate the plot of average residuals for HD\,140283 using a {\small MULTI} analysis of 51 \element{Fe} lines. We have co-added the residuals from each \element{Fe} line to smooth out any unique line defects. The residuals are of similar magnitude to those from \citet{Gallagher2010} and have similar features. Most importantly, the {\tiny MULTI} residuals for NLTE with different $\SH$ values and for LTE are all very similar. There is in fact a very slight worsening of the core fit with increased NLTE effects. It is also seen that the residuals for HD\,122563 are of a similar form, albeit with a greater magnitude. The lines analysed in HD\,122563 are far stronger than those in HD\,140283. As the lines become stronger, they become harder to fit precisely, but using NLTE profiles has no effect on improving this. 

The fact that the {\tiny MULTI} NLTE analysis has led to no improvement over the {\tiny MULTI} LTE analysis in the fit of the \element{Fe} lines is not entirely surprising. In fact when compared to the {\tiny ATLAS} analysis for the same subset of \element{Fe} lines (seen in red) there is a considerable worsening of the residuals, particularly for HD\,140283, although NLTE effects were still important to investigate. The effects of NLTE on \ion{Fe}{i} in metal-poor stars are dominated by overionisation. To a first approximation, the populations of neutral \element{Fe} energy levels can be seen as following a Boltzmann distribution relative to one another, but with a shift in the ionization equilibrium. The results show that this does not affect the shape of the profiles, though for a given equivalent width the abundance needed to reproduce the line increases. The other main NLTE effect, a change in the line and continuum opacities with height in the atmospheres, apparently has little overall effect on the \ion{Fe}{i} line profiles.

In summary, NLTE effects in \element{Fe} affect $\vconv$ by up to $\sim 0.1\,\kms$, and may affect $\fodd$ for \element{Ba} up to 0.06 or 0.09, but NLTE does not result in a \textit{better} fit to the \element{Fe} lines (Fig.~\ref{fig:NLTELTEresidualcompare}). Other mechanisms may come into play, e.g. as represented by 3D model atmospheres. It would still be interesting to see, however, how a NLTE treatment of \element{Ba} affects the inferred $\fodd$.

\section{Discussion}
\label{sec:discuss}
With the exception of one star, BD$-$04${}^{\circ}$\,3208, all of the stars analysed in this paper and HD\,140283, which was studied in detail in \citet{Gallagher2010}, show a non-physical isotope ratio ($\fodd<0.11$) close to the s-process-only composition. The non-physical results for $\fodd$ suggest that applying a 1D LTE treatment to analyse the isotopic fraction of the \element{Ba} 4554\,\AA\ line does not appear to be very robust. There are several possibilities why this might be the case, and a few possible solutions to the problem, which we now discuss. 

\subsection{Alternative broadening techniques}
\label{sec:altbroad}

We have shown that $\fodd$ has a high sensitivity to $\vconv$, which we have determined by fitting synthetic 1D LTE (and in \S\ref{sec:NLTE}, NLTE) profiles to \element{Fe} lines. There are several reasons why we chose to use \element{Fe} lines. Firstly they are the most abundant species in a metal-poor spectrum and cover a considerable range in line strength. As mentioned above, these two attributes are useful because they allow us to use lines that form at similar depths to the \element{Ba} line. However, we revisit this argument and consider whether \element{Fe} is a sensible atomic species to use to determine the macroturbulence of another species.

To test this hypothesis, we analysed several \ion{Ca}{i} lines in the star with the best quality stellar spectrum, HD\,140283 ($S/N=1100$), using the same techniques described in \S\ref{sec:LTE}. Only seven lines were found to be unblended and to have similar strengths to the \element{Ba} line ($10\leq W \ ({\rm m}\AA)\leq 50$). This meant that the standard error in $\vconv$ was much larger for this set of lines than was found for the 93 \element{Fe} lines we used in \citet{Gallagher2010}. We found that $\langle\vconv\rangle_{\element{Ca}}=5.63\pm0.13 \ \kms$. This is a smaller value than was determined using \element{Fe} lines ($5.75\pm0.02 \ \kms$) but still within the $1\sigma$ error, and the statistics of using only seven lines meant that by using \element{Ca} we would greatly increase the uncertainty in $\fodd$. Nevertheless it was found that $\fodd$, which from the \element{Fe} analysis was found to be $0.01\pm0.06$ (in the 4554\,\AA\ line\footnote{Table~\ref{tab:1Dresults} lists $\fodd$ as calculated by both the 4554 and 4934\,\AA\ lines for HD\,140283.}), would move higher to $0.10\pm0.11$. We can not say whether \element{Ca} is a better atomic species to use than \element{Fe} as the spread in $\vconv$ is too high.

To avoid using other elements to constrain the macroturbulent broadening of \element{Ba}, we experimented by treating $\vconv$ as a free parameter whilst determining $\fodd$, thus deriving it from the \element{Ba} line. This was done for all six stars in this study. Results are shown in Table~\ref{tab:barium-macro}. In Table~\ref{tab:barium-macro} we also give the standard deviation (s.d.) of the \element{Fe} line measurements, as an indication of the uncertainty associated with the measurement of just one line, e.g. \element{Ba} 4554\,\AA. As shown, $\vconv$ is little changed (within 1 s.d.) in four stars when using just the \element{Ba} line, and within 2 s.d. in a fifth star, and therefore there is little change to $\fodd$. It is interesting to note that two of the turn-off stars and the subgiant, HD\,140283, are found to have higher $\vconv$ values, driving smaller $\fodd$ (more non-physical) ratios though in all cases the differences in $\vconv$ are comparable to the uncertainty expected for one measurement. It is therefore still unclear whether the \element{Fe} analysis does describe the Doppler broadening for \element{Ba} well or not, but there is not strong evidence against it.

\begin{table}
\begin{center}
\caption{Values of $\vconv$ (measured in $\kms$) as determined from \element{Fe} lines and directly by the \ion{Ba}{ii} 4554\,\AA\ line. Note that we cannot calculate $\fodd$ in BD$+$26${}^\circ$\,3578 when we employ $\vconv \ (\element{Ba})$.}
\begin{tabular}{l c c r c r}
\hline\hline
Star		 & $\vconv$ & s.d. & $\fodd$ & $\vconv$ & $\fodd$ \\
&(\element{Fe}) &(\element{Fe})&(\element{Fe})&(\element{Ba})&(\element{Ba})	\\
\hline
HD\,122563	 &	$6.99$ & $0.54$	&	$-0.12$ & $6.82$		&	$-0.04$ \\
HD\,88609	 &	$7.03$ & $0.45$	&	$-0.02$	& $6.98$		&	$0.01$	\\
HD\,84937	 &	$6.98$ & $0.41$	&	$-0.05$	& $6.92$		&	$-0.03$	\\
BD$-$04${}^\circ$\,3208&	$6.62$ & $0.32$	&	$0.18$ 	& $6.88$		&	$-0.01$	\\
BD$+$26${}^\circ$\,3578& 	$6.41$ & $0.34$	&	$0.02$ 	& $8.85$		&	$\cdots$\\
HD\,140283	 &	$5.75$ & $0.19$	&	$0.01$ 	& $6.06$		&	$-0.23$	\\
\hline
\label{tab:barium-macro}
\end{tabular}
\end{center}
\end{table}

As well as testing whether \element{Fe} was an adequate species to use, we also considered whether a simple Gaussian adequately describes macroturbulent broadening. In \citet{Gallagher2010} we showed that a $\rt$ macroturbulent broadening mechanism better fit several \element{Fe} lines than a Gaussian. Therefore we employed a $\rt$ treatment for each star analysed in this investigation (\S\ref{sec:rt}, Fig.~\ref{fig:balinesrt}, and Table~\ref{tab:1Dresults}). It was found that for the giant stars, a radial-tangential fit the \element{Fe} lines better, with only one line in each spectrum better fit by a Gaussian. We found that for the three turn-off stars a $\rt$ technique fit $68\%-86\%$ of \element{Fe} lines better. This demonstrates a clear reason to consider using $\rt$ when working in 1D LTE over a Gaussian broadening mechanism. This is particularly clear with the giant stars, whose {Fe} line cores are closer to saturation causing the wings to become more significant. Examination of the residual plots in Fig.~\ref{fig:residplots} shows further the improvement to the fits when using $\rt$. However we stress to the reader that whilst using such a technique under the assumptions of 1D LTE seems to improve upon fitting errors associated with using a traditional Gaussian, both are still symmetric profiles and are unable to remedy the issue of asymmetries associated with absorption lines in real stellar spectra.

\subsection{Fe line residuals}
\label{sec:feresids}

\begin{figure}
\begin{center}
	\resizebox{0.99\hsize}{!}{\includegraphics{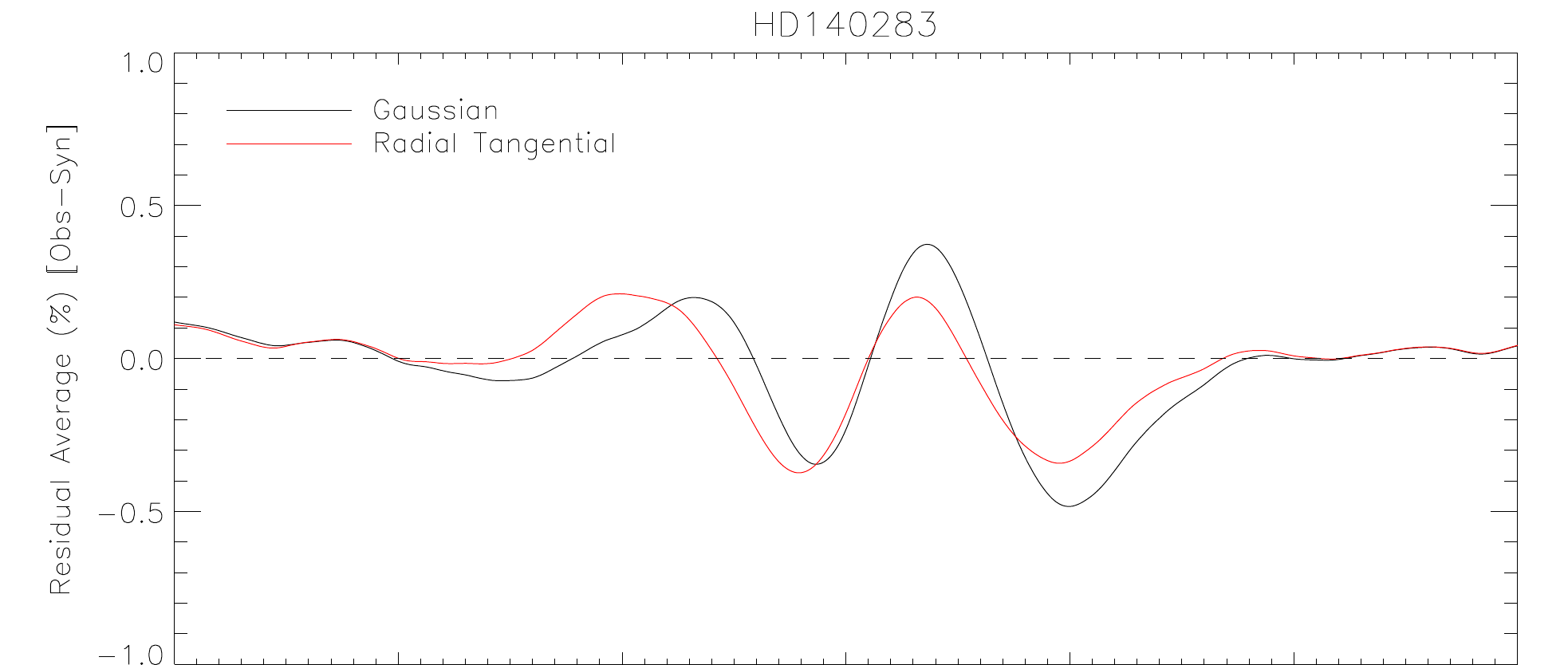}}
	\resizebox{0.99\hsize}{!}{\includegraphics{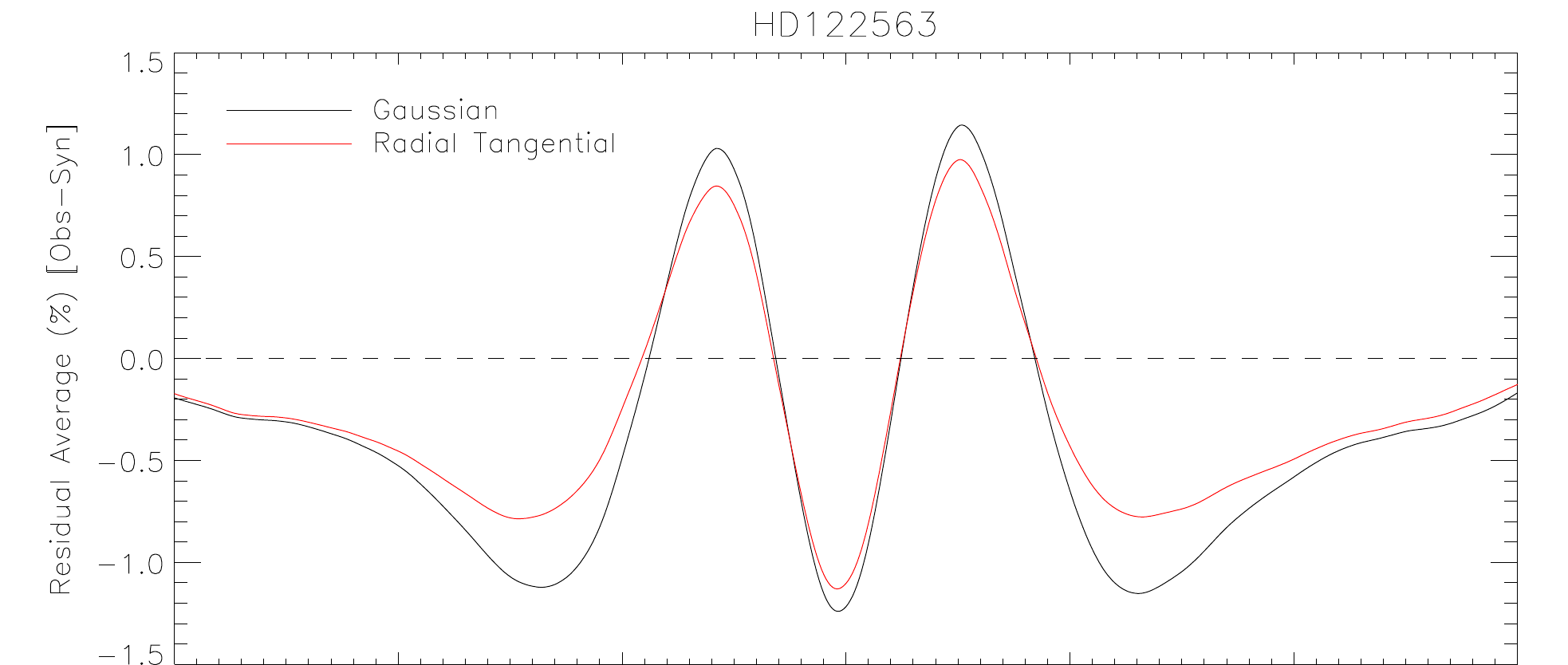}}
	\resizebox{0.99\hsize}{!}{\includegraphics{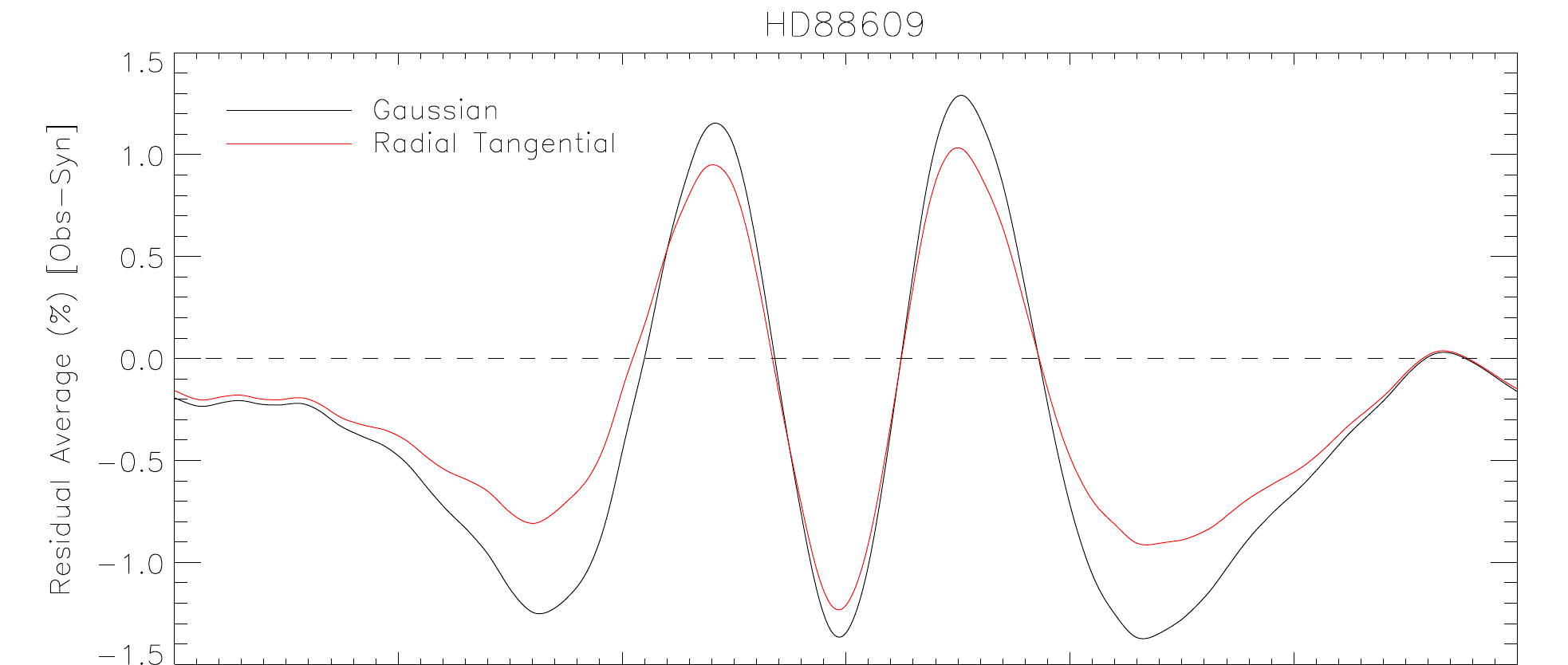}}
	\resizebox{0.99\hsize}{!}{\includegraphics{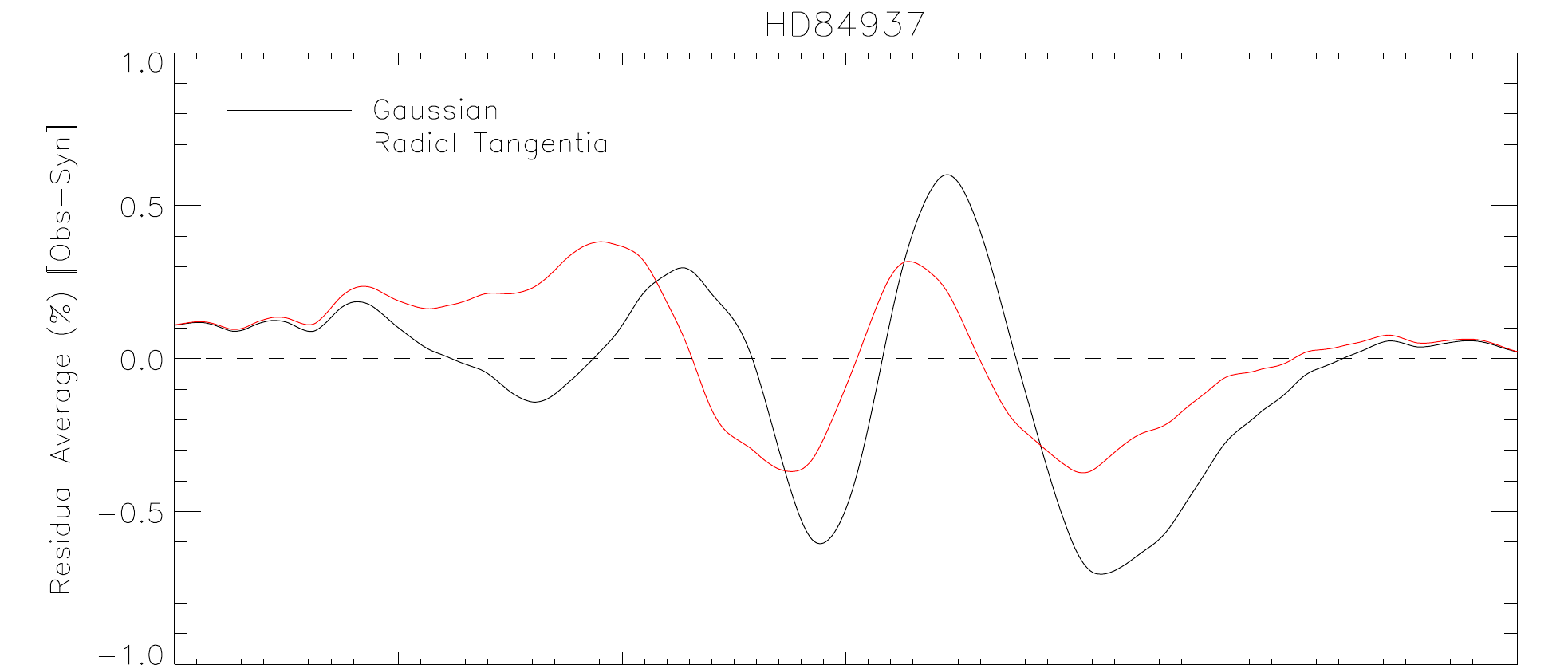}}
	\resizebox{0.99\hsize}{!}{\includegraphics{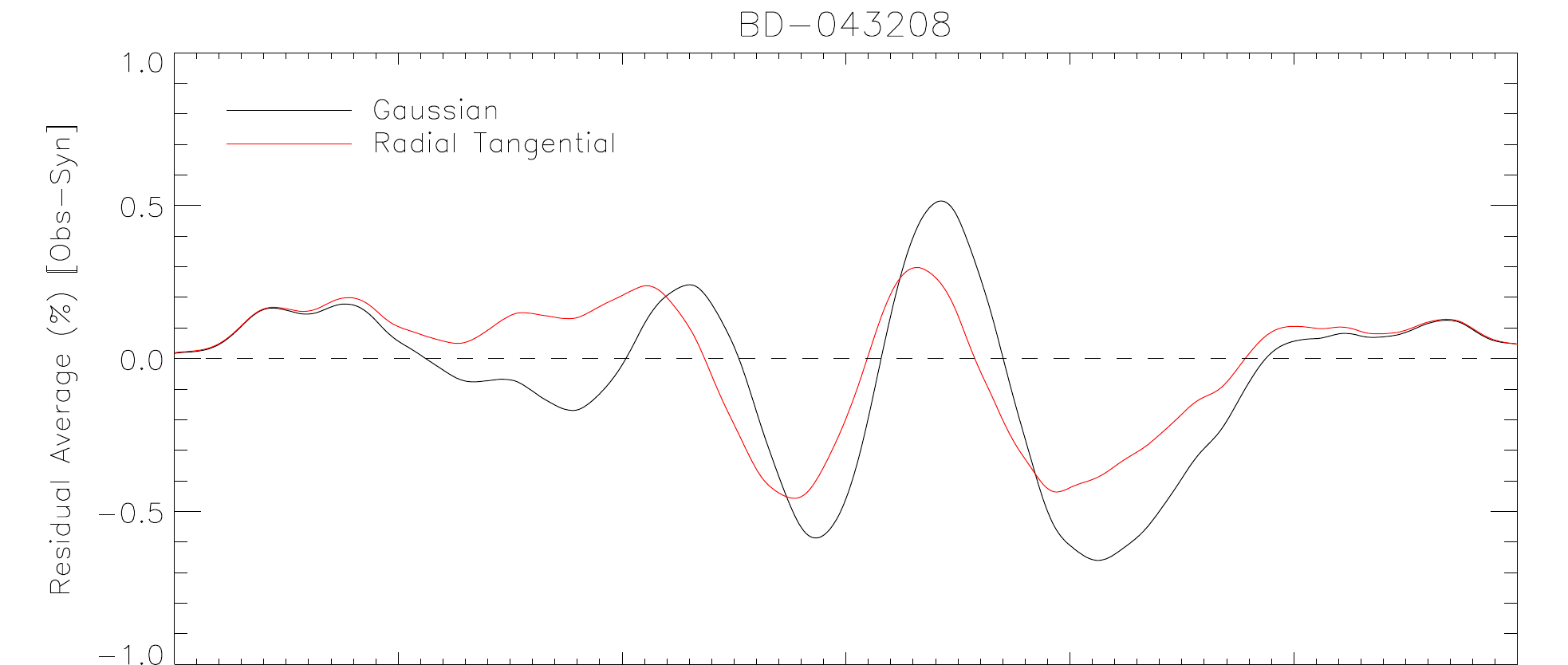}}
	\resizebox{0.99\hsize}{!}{\includegraphics{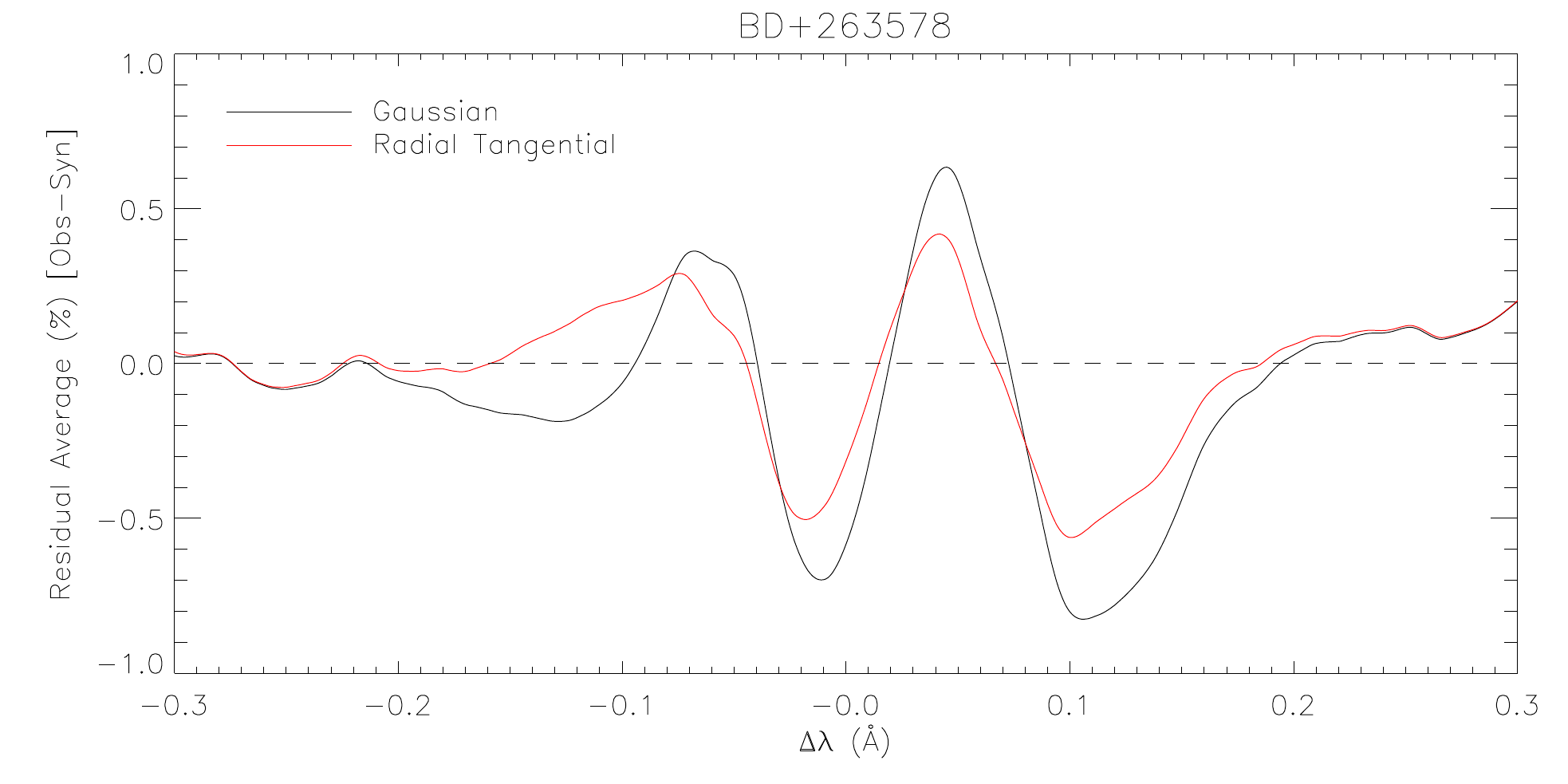}}
\caption{The average residuals of the \element{Fe} lines used to constrain $\vconv$ and $\rt$.}	
\label{fig:residplots}
\end{center}
\end{figure}

Fig.~\ref{fig:residplots} shows the average residuals of the \element{Fe} lines used in determining $\vconv$ and $\rt$. It illustrates the difficulties in fitting absorption lines with 1D LTE synthetic profiles. 

It is seen that the turn-off and subgiant stars have quite an asymmetric residual profile, with particular problems occurring in the red wings, 100 to $130\,{\rm m}\AA$ from line centre for all four stars. This seems to be caused by underlying assumptions adopted in 1D LTE radiative transfer codes; real absorption lines are not perfectly symmetric. 

This investigation is an extension of the asymmetry analysis conducted in \citet{Gallagher2010}; the asymmetry seems to occur in all four stars. Severe fitting issues occur in the giant stars, HD\,122563 and HD\,88609, where the \element{Ba} (and hence \element{Fe}) line equivalent widths are $\sim 90\,{\rm m}\AA$ (Table~\ref{tab:1Dresults}, row (27)) and lines cores begin to saturate so pressure broadening becomes more significant in the wings, and this may explain the symmetric residuals seen at $\pm0.14\,\AA$. 

\subsection{New approaches to determine isotope mixtures}
\label{sec:newapproach}
It is clear that there is still uncertainty over which atomic species to use to determine $\vconv$. However it is more likely that the assumptions that a 1D LTE code adopts, in particular the symmetric broadening mechanisms used in replicating macroscopic turbulence in a star's atmosphere, is a more likely reason for the large residuals, illustrated in Fig.~\ref{fig:residplots}. In the previous paper we asked, but did not answer, whether 3D hydrodynamical codes could solve these issues. One major drawback of using 3D techniques is the time-scales involved with the both the model atmosphere production, which currently can take several months of computation, and the radiative transfer calculations, which currently take several hours per spectral line synthesis. In addition, 3D hydrodynamics is a rather new field of analysis, and as such there has been very little published to test the reliability of the codes currently available. However it appears that 3D hydrodynamics is the next logical step to take in this field \citep{Collet2009,Ludwig2005} in response to the asymmetric profiles. However, this may not solve the problem of $\fodd$ since \citet{Collet2009} found that while the 3D model atmosphere did seem to improve upon asymmetries associated with the \element{Ba} 4554\,\AA\ absorption line, the result for $\fodd$ increased the s-process contribution relative to the 1D LTE result. Therefore, even if adopting a spectral line synthesis code that employs 3D hydrodynamic model stellar atmospheres is the next step to better understanding the problems with fitting absorption lines, it does not yet appear to be able to reduce the high s-process fractions we have found, in fact it would appear using such a technique would enhance the s-process fraction further.

In addition to new approaches in the analysis of isotopic fractions, perhaps a new, more accurate observing technique is required to get the best possible data for this type of complex analysis. Astronomers using precise radial velocities to detect extrasolar planets via the parent star Doppler ``wobble'' are employing observing procedures to attain Doppler precisions $<3\,{\rm ms}^{-1}$ by passing the starlight through an iodine (${\rm I}_2$) absorption cell. This allows them to measure very small wavelength shifts, which affect the instrumental profile \citep{Butler1996}. This can be calculated over all wavelength scales and in each short exposure, to determine the instrumental broadening profile at any time. 

\citet{Aoki2004} have shown that there is a very slight asymmetry in the \element{ThAr} lines recorded with HDS, at least in the HD\,140283 observing run, which were used to calculate $\nu_{\rm inst}$. Using the ${\rm I}_2$ method described above, it is possible to deconvolve the instrumental profile from the stellar spectrum, leaving only real stellar broadening in the lines. This would remove most of the uncertainty surrounding the instrumental profile from any subsequent analysis. Issues with this method are the requirement for a stable ${\rm I}_2$ cell and the large amount of time that would be required to complete such a complex reduction. In conjunction with a 3D hydrodynamical treatment this might require a lot of time and still may not explain the r- and s-process contributions any better than the method described in this paper. However, it is a method still favoured by astronomers looking for radial velocity fluctuations in stars, as it yields the smallest error in the spectral data. The difficulty in modelling the \element{Fe} lines, let alone \element{Ba}, may justify the additional effort required.

\section{Conclusions}
\label{sec:conclusion}
We have carried out a careful examination of the \element{Ba} isotopic ratios for five metal-poor stars using a 1D LTE treatment in conjunction with high resolution ($R\equiv\lambda/\Delta\lambda\geq 90\,000$), high signal-to-noise ($S/N\geq 550$) spectra. We found that all stars show a high s-process signature, and only BD$-$04${}^{\circ}$\,3208 had an $\fodd$ value ($0.18\pm0.08$) that was physical, $0.11\geq\fodd\geq0.46$ according to the isotopic abundance determinations by \citet{Arlandini1999}. According to these limits, all other stars yield a non-physical isotopic fraction, $\fodd<0.11$.

Using the radiative transfer code {\tiny MULTI} with {\tiny MARCS} 1D model atmospheres, we have also applied a 1D NLTE treatment to \element{Fe} lines in HD\,122563 and HD\,140283. We found that using NLTE did not fit the \element{Fe} line cores better, rather it increased the core residual in both stars, see Fig.~\ref{fig:NLTELTEresidualcompare}. A large NLTE effect ($\SH=0.001$) did, however, reduce $\vconv$ by $0.1\,\kms$, and may imply an increase of $\fodd$ by $\sim 0.06-0.09$, which would reduce the non-physicality of the results.

During the investigation we have shown how 1D LTE radiative transfer codes and model atmospheres seem to be inadequate to determine $\fodd$. In particular we have shown that asymmetries in the red wings are observed in the \element{Fe} line residuals for all four turn-off and subgiant stars, while the two giants have larger, albeit symmetric, residuals. We speculate that spectral line synthesis with 3D hydrodynamic model stellar atmospheres would improve residuals in both the \element{Fe} and \element{Ba} lines, but we make no predictions as to whether it would improve upon the obvious problems with non-physical isotopic fractions we have encountered here. In addition we have asked, but not answered, whether taking the observations using an ${\rm I}_2$ cell, designed to improve accuracy and quality of the data, would help to better constrain $\fodd$. Certainly if one compares the error estimation of $\fodd$ in HD\,140283 ($\pm0.06$, $S/N=1100$) with HD\,84937 ($\pm0.11$, $S/N=630$), there is almost double the uncertainty in HD\,84937 than in HD\,140283. We speculate that this could be due to the quality of the spectral data used, however, we acknowledge that this could also be a result of the difference in the number of \element{Fe} lines used to determine $\vconv$, as we measure error by the standard error. However, $\sigma_{\vconv , {\rm HD\,84937}} =0.06\,\kms$ is fairly typical as an error estimate for these stars, yet $\sigma_{\fodd}$ is larger in HD\,84937 than in all other stars.

It was found that using a radial-tangential broadening technique rather than a traditional Gaussian improved upon fitting errors for both the \element{Fe} and \element{Ba} lines. Whilst using such a technique does not remove asymmetries seen in absorption lines for real stellar data, it appears that if one were to adopt a 1D LTE approach to resolve an absorption line, a radial-tangential profile should be used to model the macroturbulence of the star rather than a Gaussian, although a Gaussian could still be used to model the instrumental profile.

We have also conducted a study of \element{Eu} abundances in the two giants, HD\,122563 and HD\,88609. This was used to determine [Ba/Eu] ratios, which are used to assess the star's s- and r-process content. Both stars have [Ba/Eu] ratios that indicate a large r-process contribution, which agrees well with \citet{Honda2006} \& \citet{Honda2007}, but contradicts the isotope analysis conducted in this paper. This further justifies scepticism of the 1D LTE techniques employed in our \element{Ba} isotope analysis.

We have increased the number of stars that have undergone a \element{Ba} isotope analysis. However it is difficult to believe that all stars analysed have a significant s-process contribution, or s-process enhancement in contradiction of the (limited) [Ba/Eu] data and the strengths of the \citet{Truran1981} theory and \citet{Travaglio1999} calculations. It is much more likely that the symmetric 1D LTE techniques used in this investigation are inadequate and improvements to isotopic ratio analysis need to be made.

\nocite{*}
\bibliographystyle{aa}
\bibliography{18382}

\begin{thebibliography}{50}
\expandafter\ifx\csname natexlab\endcsname\relax\def\natexlab#1{#1}\fi

\bibitem[{{Anders} \& {Grevesse}(1989)}]{Anders1989}
{Anders}, E. \& {Grevesse}, N. 1989, \gca, 53, 197

\bibitem[{{Aoki} {et~al.}(2009){Aoki}, {Barklem}, {Beers}, {Christlieb},
  {Inoue}, {Garc{\'{\i}}a P{\'e}rez}, {Norris}, \& {Carollo}}]{Aoki2009}
{Aoki}, W., {Barklem}, P.~S., {Beers}, T.~C., {et~al.} 2009, \apj, 698, 1803

\bibitem[{{Aoki} {et~al.}(2004){Aoki}, {Inoue}, {Kawanomoto}, {Ryan}, {Smith},
  {Suzuki}, \& {Takada-Hidai}}]{Aoki2004}
{Aoki}, W., {Inoue}, S., {Kawanomoto}, S., {et~al.} 2004, \aap, 428, 579

\bibitem[{{Arlandini} {et~al.}(1999){Arlandini}, {K{\"a}ppeler}, {Wisshak},
  {Gallino}, {Lugaro}, {Busso}, \& {Straniero}}]{Arlandini1999}
{Arlandini}, C., {K{\"a}ppeler}, F., {Wisshak}, K., {et~al.} 1999, \apj, 525,
  886

\bibitem[{{Becker} {et~al.}(1993){Becker}, {Enders}, {Werth}, \&
  {Dembczynski}}]{Becker1993}
{Becker}, O., {Enders}, K., {Werth}, G., \& {Dembczynski}, J. 1993, \pra, 48,
  3546

\bibitem[{{Biemont} {et~al.}(1982){Biemont}, {Karner}, {Meyer}, {Traeger}, \&
  {Zu Putlitz}}]{Biemont1982}
{Biemont}, E., {Karner}, C., {Meyer}, G., {Traeger}, F., \& {Zu Putlitz}, G.
  1982, \aap, 107, 166

\bibitem[{{Bisnovatyi-Kogan} \& {Chechetkin}(1979)}]{Bisnovatyi1979}
{Bisnovatyi-Kogan}, G.~S. \& {Chechetkin}, V.~M. 1979, Soviet Physics Uspekhi,
  127, 263

\bibitem[{{Burbidge} {et~al.}(1957){Burbidge}, {Burbidge}, {Fowler}, \&
  {Hoyle}}]{B2FH1957}
{Burbidge}, E.~M., {Burbidge}, G.~R., {Fowler}, W.~A., \& {Hoyle}, F. 1957,
  Reviews of Modern Physics, 29, 547

\bibitem[{{Burris} {et~al.}(2000){Burris}, {Pilachowski}, {Armandroff},
  {Sneden}, {Cowan}, \& {Roe}}]{Burris2000}
{Burris}, D.~L., {Pilachowski}, C.~A., {Armandroff}, T.~E., {et~al.} 2000,
  \apj, 544, 302

\bibitem[{{Butler} {et~al.}(1996){Butler}, {Marcy}, {Williams}, {McCarthy},
  {Dosanjh}, \& {Vogt}}]{Butler1996}
{Butler}, R.~P., {Marcy}, G.~W., {Williams}, E., {et~al.} 1996, \pasp, 108, 500

\bibitem[{{Carlsson}(1986)}]{Carlsson1986}
{Carlsson}, M. 1986, Uppsala Astronomical Observatory Reports, 33

\bibitem[{{Collet} {et~al.}(2009){Collet}, {Asplund}, \& {Nissen}}]{Collet2009}
{Collet}, R., {Asplund}, M., \& {Nissen}, P.~E. 2009, Publications of the
  Astronomical Society of Australia, 26, 330

\bibitem[{{Collet} {et~al.}(2005){Collet}, {Asplund}, \&
  {Th{\'e}venin}}]{Collet2005}
{Collet}, R., {Asplund}, M., \& {Th{\'e}venin}, F. 2005, \aap, 442, 643

\bibitem[{{Cottrell} \& {Norris}(1978)}]{Cottrell1978}
{Cottrell}, P.~L. \& {Norris}, J. 1978, \apj, 221, 893

\bibitem[{{Drawin}(1968)}]{Drawin1968}
{Drawin}, H.-W. 1968, Zeitschrift fur Physik, 211, 404

\bibitem[{{Drawin}(1969)}]{Drawin1969}
{Drawin}, H.~W. 1969, Zeitschrift fur Physik, 225, 483

\bibitem[{{Fran{\c c}ois} {et~al.}(2007){Fran{\c c}ois}, {Depagne}, {Hill},
  {Spite}, {Spite}, {Plez}, {Beers}, {Andersen}, {James}, {Barbuy}, {Cayrel},
  {Bonifacio}, {Molaro}, {Nordstr{\"o}m}, \& {Primas}}]{Francois2007}
{Fran{\c c}ois}, P., {Depagne}, E., {Hill}, V., {et~al.} 2007, \aap, 476, 935

\bibitem[{{Gallagher} {et~al.}(2010){Gallagher}, {Ryan}, {Garc{\'{\i}}a
  P{\'e}rez}, \& {Aoki}}]{Gallagher2010}
{Gallagher}, A.~J., {Ryan}, S.~G., {Garc{\'{\i}}a P{\'e}rez}, A.~E., \& {Aoki},
  W. 2010, \aap, 523, A24

\bibitem[{{Garc{\'{\i}}a P{\'e}rez} {et~al.}(2009){Garc{\'{\i}}a P{\'e}rez},
  {Aoki}, {Inoue}, {Ryan}, {Suzuki}, \& {Chiba}}]{Garcia2009}
{Garc{\'{\i}}a P{\'e}rez}, A.~E., {Aoki}, W., {Inoue}, S., {et~al.} 2009, \aap,
  504, 213

\bibitem[{{Gray}(2008)}]{Graybook}
{Gray}, D.~F. 2008, {The Observation and Analysis of Stellar Photospheres}, ed.
  {Gray, D.~F.}

\bibitem[{{Honda} {et~al.}(2007){Honda}, {Aoki}, {Ishimaru}, \&
  {Wanajo}}]{Honda2007}
{Honda}, S., {Aoki}, W., {Ishimaru}, Y., \& {Wanajo}, S. 2007, \apj, 666, 1189

\bibitem[{{Honda} {et~al.}(2006){Honda}, {Aoki}, {Ishimaru}, {Wanajo}, \&
  {Ryan}}]{Honda2006}
{Honda}, S., {Aoki}, W., {Ishimaru}, Y., {Wanajo}, S., \& {Ryan}, S.~G. 2006,
  \apj, 643, 1180

\bibitem[{Hosford(2010)}]{Adamthesis}
Hosford, A. 2010, PhD thesis, University of Hertfordshire

\bibitem[{{Hosford} {et~al.}(2010){Hosford}, {Garc{\'{\i}}a P{\'e}rez},
  {Collet}, {Ryan}, {Norris}, \& {Olive}}]{Hosford2010}
{Hosford}, A., {Garc{\'{\i}}a P{\'e}rez}, A.~E., {Collet}, R., {et~al.} 2010,
  \aap, 511, A47

\bibitem[{{Imshennik}(1992)}]{Imshennik1992}
{Imshennik}, V.~S. 1992, Soviet Astronomy Letters, 18, 194

\bibitem[{{Korn} {et~al.}(2003){Korn}, {Shi}, \& {Gehren}}]{Korn2003}
{Korn}, A.~J., {Shi}, J., \& {Gehren}, T. 2003, \aap, 407, 691

\bibitem[{{Krebs} \& {Winkler}(1960)}]{Krebs1960}
{Krebs}, K. \& {Winkler}, R. 1960, Zeitschrift fur Physik, 160, 320

\bibitem[{{Lambert} \& {Allende Prieto}(2002)}]{Lambert2002}
{Lambert}, D.~L. \& {Allende Prieto}, C. 2002, \mnras, 335, 325

\bibitem[{{Ludwig} \& {Ku{\v c}inskas}(2005)}]{Ludwig2005}
{Ludwig}, H.-G. \& {Ku{\v c}inskas}, A. 2005, in ESA Special Publication, Vol.
  560, 13th Cambridge Workshop on Cool Stars, Stellar Systems and the Sun, ed.
  {F.~Favata, G.~A.~J.~Hussain, \& B.~Battrick}, 319--+

\bibitem[{{Magain}(1995)}]{Magain1995}
{Magain}, P. 1995, \aap, 297, 686

\bibitem[{{Mashonkina} {et~al.}(2010){Mashonkina}, {Gehren}, {Shi}, {Korn}, \&
  {Grupp}}]{Mashonkina2010}
{Mashonkina}, L., {Gehren}, T., {Shi}, J., {Korn}, A., \& {Grupp}, F. 2010, in
  IAU Symposium, Vol. 265, IAU Symposium, ed. {K.~Cunha, M.~Spite, \&
  B.~Barbuy}, 197--200

\bibitem[{{Mashonkina} {et~al.}(2003){Mashonkina}, {Gehren}, {Travaglio}, \&
  {Borkova}}]{Mashonkina2003}
{Mashonkina}, L., {Gehren}, T., {Travaglio}, C., \& {Borkova}, T. 2003, \aap,
  397, 275

\bibitem[{{Mashonkina} {et~al.}(2008){Mashonkina}, {Zhao}, {Gehren}, {Aoki},
  {Bergemann}, {Noguchi}, {Shi}, {Takada-Hidai}, \& {Zhang}}]{Mashonkina2008}
{Mashonkina}, L., {Zhao}, G., {Gehren}, T., {et~al.} 2008, \aap, 478, 529

\bibitem[{{Noguchi} {et~al.}(2002){Noguchi}, {Aoki}, {Kawanomoto}, {Ando},
  {Honda}, {Izumiura}, {Kambe}, {Okita}, {Sadakane}, {Sato}, {Tajitsu},
  {Takada-Hidai}, {Tanaka}, {Watanabe}, \& {Yoshida}}]{Noguchi2002}
{Noguchi}, K., {Aoki}, W., {Kawanomoto}, S., {et~al.} 2002, \pasj, 54, 855

\bibitem[{{Panov} \& {Janka}(2009)}]{Panov2009}
{Panov}, I.~V. \& {Janka}, H.-T. 2009, \aap, 494, 829

\bibitem[{{Rutten}(1978)}]{Rutten1978}
{Rutten}, R.~J. 1978, \solphys, 56, 237

\bibitem[{{Shchukina} {et~al.}(2005){Shchukina}, {Trujillo Bueno}, \&
  {Asplund}}]{Shchukina2005}
{Shchukina}, N.~G., {Trujillo Bueno}, J., \& {Asplund}, M. 2005, \apj, 618, 939

\bibitem[{{Smith} \& {Lambert}(1988)}]{Smith1988}
{Smith}, V.~V. \& {Lambert}, D.~L. 1988, \apj, 333, 219

\bibitem[{{Sneden} {et~al.}(2008){Sneden}, {Cowan}, \& {Gallino}}]{Sneden2008}
{Sneden}, C., {Cowan}, J.~J., \& {Gallino}, R. 2008, \araa, 46, 241

\bibitem[{{Spite} \& {Spite}(1978)}]{Spite1978}
{Spite}, M. \& {Spite}, F. 1978, \aap, 67, 23

\bibitem[{{Straniero} {et~al.}(1997){Straniero}, {Chieffi}, {Limongi}, {Busso},
  {Gallino}, \& {Arlandini}}]{Straniero1997}
{Straniero}, O., {Chieffi}, A., {Limongi}, M., {et~al.} 1997, \apj, 478, 332

\bibitem[{{Tajitsu} {et~al.}(2010){Tajitsu}, {Aoki}, \& {Narita}}]{Tajitsu2010}
{Tajitsu}, A., {Aoki}, W.~{Kawanomoto}, S., \& {Narita}, N. 2010, Publications
  of the National Astronomical Observatory of Japan, 13, 1

\bibitem[{{Th{\'e}venin} \& {Idiart}(1999)}]{Thevenin1999}
{Th{\'e}venin}, F. \& {Idiart}, T.~P. 1999, \apj, 521, 753

\bibitem[{{Travaglio} {et~al.}(1999){Travaglio}, {Galli}, {Gallino}, {Busso},
  {Ferrini}, \& {Straniero}}]{Travaglio1999}
{Travaglio}, C., {Galli}, D., {Gallino}, R., {et~al.} 1999, \apj, 521, 691

\bibitem[{{Truran}(1981)}]{Truran1981}
{Truran}, J.~W. 1981, \aap, 97, 391

\bibitem[{{Villemoes} {et~al.}(1993){Villemoes}, {Arnesen}, {Heijkenskjold}, \&
  {Wannstrom}}]{Villemoes1993}
{Villemoes}, P., {Arnesen}, A., {Heijkenskjold}, F., \& {Wannstrom}, A. 1993,
  Journal of Physics B Atomic Molecular Physics, 26, 4289

\bibitem[{{Wanajo} \& {Ishimaru}(2006)}]{Wanajo2006}
{Wanajo}, S. \& {Ishimaru}, Y. 2006, Nuclear Physics A, 777, 676

\bibitem[{{Wanajo} {et~al.}(2001){Wanajo}, {Kajino}, {Mathews}, \&
  {Otsuki}}]{Wanajo2001}
{Wanajo}, S., {Kajino}, T., {Mathews}, G.~J., \& {Otsuki}, K. 2001, \apj, 554,
  578

\bibitem[{{Wanajo} {et~al.}(2003){Wanajo}, {Tamamura}, {Itoh}, {Nomoto},
  {Ishimaru}, {Beers}, \& {Nozawa}}]{Wanajo2003}
{Wanajo}, S., {Tamamura}, M., {Itoh}, N., {et~al.} 2003, \apj, 593, 968

\bibitem[{{Wendt} {et~al.}(1984){Wendt}, {Ahmad}, {Buchinger}, {Mueller},
  {Neugart}, \& {Otten}}]{Wendt1984}
{Wendt}, K., {Ahmad}, S.~A., {Buchinger}, F., {et~al.} 1984, Zeitschrift fur
  Physik, 318, 125

\end{thebibliography}
\end{document}